\documentclass{article}

\usepackage{amsmath}
\usepackage{natbib}
\usepackage[title]{appendix}
\usepackage[margin=1.0in]{geometry}
\usepackage{times}

\numberwithin{equation}{section}

\title{\bf Non-trivial and trivial conservation laws in covariant formulations of geophysical fluid dynamics}

\author{Martin Charron and Ayrton Zadra \\
	Recherche en pr\'evision num\'erique atmosph\'erique \\
	Environnement et Changement climatique Canada, Dorval, Qc, Canada}

\date{\today}

\begin{document}

\maketitle

\begin{abstract}
Using a manifestly invariant Lagrangian density based on Clebsch fields and suitable for geophysical fluid dynamics, the conservation of mass, entropy, momentum and energy, and the associated symmetries are investigated. In contrast, it is shown that the conservation of Ertel's potential vorticity is not associated with any symmetry of the equations of motion, but is instead a trivial conservation law of the second kind. This is at odds with previous studies which claimed that potential vorticity conservation relates to a symmetry under particle-relabeling transformations. From the invariant Lagrangian density, a canonical Hamiltonian formulation is obtained in which Dirac constraints explicitly include the (possibly time-dependent) metric tensor. In this case, it is shown that all Dirac constraints are primary and of the second class, which implies that no local gauge symmetry transformations of Clebsch fields exist. Finally, the corresponding non-canonical Hamiltonian structure with time-dependent strong constraints is derived using tensor components. The existence of Casimir invariants is then investigated in arbitrary coordinates for two choices of dynamical variables in phase space.
\end{abstract}

\section{Introduction}
Hamilton's least action principle provides an elegant way of describing the dynamics of physical systems. Equations of motion arise from a variational problem for the action functional, which on its turn is defined by a single fundamental object called Lagrangian density. In this case, preservation of physical properties --- e.g.\ coordinate-choice independence (i.e.\ covariance), the adequate approximate limits, the underlying symmetries and corresponding conservation laws, and so on --- translate into constraints on the form of the acceptable Lagrangian densities.

To render covariance explicit, all results presented in this paper are written in terms of tensor components suitable for any non-inertial coordinate systems in which time intervals are absolute. This ensures that the underlying theory is relativistic \citep{Charron15b}, obviously not in the traditional sense of describing fluids with velocities comparable to that of light, but in the literal sense of ``obeying a principle of relativity'' --- in this case, the principle of Newtonian relativity. Using Dirac's theory for constrained systems \citep{Dirac50,Dirac64}, the corresponding Hamiltonian formulation is also presented. This may be a novel perspective on Hamiltonian dynamics applied to fluid motion for which coordinates are as general as they can be in the context of Newtonian mechanics. In order to provide a consistent overview of the topic, some developments presented here are simply a re-derivation in tensor form or a generalization of known results, in which case the appropriate citations are provided.

Symmetries and conservation laws are studied and Noether's theorems are revisited in the context of a covariant formulation of fluid mechanics, implying that results presented here are valid in all admissible coordinate systems. In particular, Noether's second theorem is also discussed due to its relation to local gauge invariances in under-determined dynamical systems and because it is sometimes cited by authors associating Ertel's theorem with a symmetry. The Hamiltonian formalism with constraints is also used since it provides a powerful tool to verify whether Noether's second theorem is relevant in the study of fluid dynamics.

Previous work by \citet{Newcomb67,Bretherton70,Ripa81,Salmon82,Salmon88,Salmon98,Salmon13,Muller95,Padhye96} associated Ertel's theorem (the conservation of potential vorticity for inviscid fluids) with a particular symmetry of the action functional and/or Lagrangian density in label representation. Here, it is argued that this association is unfounded.

The formalism and the Lagrangian density used in this paper are outlined in section \ref{overview}. In section \ref{sym}, a demonstration of Noether's first theorem is presented. The demonstration considers field transformations as well as coordinate transformations, and accounts for the presence of the metric tensor and external forcing. Symmetries of the equations of motion leading to mass, entropy, momentum, and energy conservation are studied in the context of arbitrary coordinates. In section \ref{secertel}, the relabeling transformation of particle-like formulations is analyzed in the context of a covariant field theory. It is demonstrated that particle-relabeling is not associated with Ertel's theorem and that Ertel's potential vorticity conservation is in fact a trivial law of the second kind, following the terminology of \citet{Olver93}. In section \ref{ham}, canonical and non-canonical Hamiltonian formulations are presented in arbitrary coordinates. It is shown that the constraints on momenta and fields, and the non-vanishing determinant of their corresponding Poisson brackets, imply that the equations of motion for inviscid fluid dynamics (unapproximated and geometrically approximated) are not invariant under local gauge transformations. Noether's second theorem is therefore of no relevance in that case. It is also outlined that a trivial conservation law of the second kind in fluid mechanics --- that is, Ertel's theorem --- translates into a Casimir invariant in a non-canonical Hamiltonian formulation. Conclusions are drawn in section \ref{conclu}.

\section{A manifestly invariant Lagrangian density} \label{overview}
In this section, the main results from \citet{Zadra15} are reviewed and summarized. A Lagrangian density ${\cal L}$ is a scalar that can be used to describe the evolution of a dynamical system. Hamilton's least action principle states that solutions of the equations of motion are those whose action functional
\begin{align}
{\cal S} = \int d^4x \sqrt{g} {\cal L} \label{actionfun}
\end{align}
is an extremum, i.e.\ when
\begin{align}
\delta {\cal S}=\int d^4x \sqrt{g} \sum_{p=1}^{P} \Lambda_{(\psi_{(p)})} \delta\psi_{(p)} =0,
\end{align}
where $P$ is the total number of dynamical fields $\psi_{(p)}$, and $\delta\psi_{(p)}$ indicates arbitrary perturbations of those fields with given and fixed initial, final, and boundary values. The quantity $d^4x \sqrt{g} \equiv dx^0dx^1dx^2dx^3 \sqrt{g}$, where $x^0$ is time and $x^1$, $x^2$, $x^3$ are spatial curvilinear coordinates, is an invariant space-time volume element, and $g$ is the determinant of the symmetric covariant metric tensor $g_{\mu\nu}$. This condition on the action leads to the Euler-Lagrange equations. In the case where the Lagrangian density ${\cal L}$ may be solely written in terms of scalar fields and their first derivatives, and assuming the metric tensor is prescribed and not a dynamical field (as is the case in geophysical fluid dynamics), the Euler-Lagrange equations of motion take the form
\begin{align}
\Lambda_{(\psi_{(p)})}=0, \label{m0}
\end{align}
where
\begin{align}
\Lambda_{(\psi_{(p)})}=\frac{\partial {\cal L}}{\partial \psi_{(p)}} - \left( \frac{\partial {\cal L}}{\partial (\psi_{(p),\mu})} \right)_{:\mu} \label{motion}
\end{align}
for all the scalar fields on which the Lagrangian density depends \citep{Zadra15}. The formalism and notation used in this paper are those of \citet{Charron14a, Charron14b, Charron15a, Zadra15}. In particular, tensors are defined with respect to synchronous coordinate transformations (i.e.\ time intervals are absolute). The admissible spatial coordinate transformations are otherwise general, and may involve the time. A comma followed by an index (here $\mu$) indicates an ordinary partial derivative with respect to $x^\mu$, where $\mu$ takes on the values $0$, $1$, $2$, or $3$. A colon followed by an index indicates a covariant derivative. Repeated Greek indices are summed from $0$ to $3$, and Latin indices from $1$ to $3$, unless otherwise indicated.

\citet{Zadra15} have shown that the manifestly invariant Lagrangian density
\begin{align}
{\cal L}=-\rho \left( \frac{1}{2}h^{\mu\nu}v_\mu v_\nu+\Phi+I+g^{0\mu}v_\mu \right) \label{lag}
\end{align}
suitably describes the Newtonian dynamics of an inviscid fluid in arbitrary coordinates under the influence of an external scalar gravitational potential $\Phi$. The scalar internal energy of the fluid is described by $I=I(\rho,s)$, where $\rho$ is the fluid density, and $s$ its specific entropy. The covariant 4-vector field $v_\mu$ is given by a Clebsch decomposition,
\begin{align}
v_\mu = \alpha_{,\mu} + \beta s_{,\mu} + \sum_{r=1}^N \gamma_{(r)} \lambda_{(r),\mu}, \label{vvv}
\end{align}
where the scalar fields $\alpha$, $\beta$, $\gamma_{(r)}$, and $\lambda_{(r)}$ are Clebsch potentials, with $N \ge 2$, suitable for any flow \citep{Bretherton70}. Note that the specific entropy scalar field $s$ is also interpreted as a Clebsch potential. Taking into account the density and the ensemble of Clebsch potentials, there are $P=4+2N$ dynamical scalar fields.

The symmetric contravariant tensor $h^{\mu\nu}=g^{\mu\nu}-g^{0\mu}g^{0\nu}$, written in terms of the symmetric contravariant space-time metric tensor $g^{\mu\nu}$, describes the geometry of space (instead of space-time), within the admissible class of synchronous coordinate transformations. The term $\rho h^{\mu\nu}v_\mu v_\nu/2$ is the scalar kinetic energy density of the flow as calculated in an inertial coordinate system. Because this kinetic energy density is a manifestly invariant field, its value at any given point in space-time is the same in all admissible (inertial as well as non-inertial) coordinate systems. The scalar $g^{0\mu}v_\mu$ includes the contribution of non-inertial acceleration terms to the Lagrangian density. The absolute nature of time intervals in Newtonian mechanics imposes a constraint on the space-time metric tensor: the component $g^{00}$ must be a non-zero constant (taken here as unity). Therefore, the contravariant 4-vector $h^{0\mu}=h^{\mu 0}$ identically vanishes. The contravariant 4-velocity vector field $u^\mu=dx^\mu/dt$ is taken to have component $u^0=dx^0/dt=1$. It is related to the covariant 4-vector field $v_\mu$ by
\begin{align}
u^\mu \equiv h^{\mu\nu} v_\nu + g^{0\mu}. \label{vtou}
\end{align}

The above Lagrangian density and Euler-Lagrange equations lead to the equations governing the evolution of the Clebsch fields and fluid density, which are provided by eqs (65)--(70) in \citet{Zadra15}. The momentum equations are also derived from these equations in their Appendix E.

The fact that the above Lagrangian density is manifestly invariant (i.e.\ written with tensors only) ensures that any approximation obtained from it which preserves its scalar property will remain consistent, to a chosen order of approximation, with the laws of classical physics. \citet{Staniforth14, Charron15a, Zadra15} and many others provide examples of such an approach (with or without using an action functional). Approximated equations of motion must be covariant, although that does not imply that any covariant equations are dynamically consistent equations of motion in classical physics. An order by order consistency with the laws of classical physics is also required.

The above description of fluid dynamics involves a Clebsch decomposition of the 4-velocity field. In practice, given a vector field, Clebsch potentials may be difficult to obtain. A particle-like Lagrangian formulation without Clebsch potentials exists at the expense of imposing {\it a priori} constraints on mass and entropy conservation. See \citet{Salmon98,Zadra15} and references therein. A field formulation in arbitrary coordinates is arguably more satisfying, as mass and entropy conservation equations emerge from Hamilton's least action principle as dynamical equations. In the following, both formulations will be considered.

\section{Noether's first theorem and non-trivially conserved 4-currents}\label{sym}
Since \citet{Noether18} and \citet{Bessel-Hagen21}, it is known that continuous symmetries of the equations of motion lead to conserved currents. For covariant theories defined in 4-dimensional space-time, these conservation laws may be written as 4-dimensional divergent-free equations, e.g.\ ${j^\mu}_{:\mu}=0$ or equivalently $(\sqrt{g} \, j^\mu)_{,\mu}=0$. In this section, Noether's first theorem is re-derived in the specific context of the formalism outlined in section \ref{overview}: synchronous coordinate transformations in a metric space-time \citep[not necessarily flat, thus allowing for geometric approximations, see][]{Charron14b} with a Lagrangian density depending on scalar fields and their first derivatives, the metric tensor, and possibly an external potential.

The symmetry transformations here considered are generated by: 1) transformations of the dynamical fields themselves at a given point in space-time, called {\it active} field transformations; and 2) transformations induced by changes of the coordinates --- which may be interpreted as {\it passive} transformations at a given point in space-time, as will be seen below.

\subsection{Active transformations}
First, consider infinitesimal active field transformations only. Perturbations to the scalar fields $\psi_{(p)}(x)=\psi_{(p)}(x^0,x^1,x^2,x^3)$ are written as
\begin{align}
\bar{\delta} \psi_{(p)} \equiv \bar{\psi}_{(p)}(x) - \psi_{(p)}(x), \label{commbar}
\end{align}
where $\bar{\psi}_{(p)}$ is a perturbed scalar field. The perturbations $\bar{\delta} \psi_{(p)}$ may depend on all the fields, but they take place at fixed space-time points. They leave the metric tensor, its determinant, and all external fields unchanged. It follows that active transformations and gradients are operations that commute, i.e.\
\begin{align}
\bar{\delta} (\psi_{(p),\mu}) = (\bar{\delta} \psi_{(p)})_{,\mu}. \label{sw}
\end{align}

Any active transformation $\bar{\delta}\psi_{(p)}$ induces a perturbation to the integrand of the action functional, which may be written as
\begin{align}
\bar{\delta}(\sqrt{g}{\cal L})&=\sqrt{g}\;\bar{\delta}{\cal L}, \nonumber \\ 
&=\sqrt{g} \sum_{p=1}^{P} \left( \frac{\partial {\cal L}}{\partial \psi_{(p)}} \bar{\delta} \psi_{(p)} + \frac{\partial {\cal L}}{\partial (\psi_{(p),\mu})} \bar{\delta} (\psi_{(p),\mu}) \right), \nonumber \\
&=\sqrt{g} \sum_{p=1}^{P} \Lambda_{(\psi_{(p)})} \bar{\delta} \psi_{(p)} +\sqrt{g} \left( \sum_{p=1}^{P} \frac{\partial {\cal L}}{\partial (\psi_{(p),\mu})} \bar{\delta} \psi_{(p)}\right)_{:\mu}
\end{align}
from Eqs \eqref{motion} and \eqref{sw}. The relation above is true, whether the transformation is a symmetry or not.

If in addition one finds that a particular transformation $\bar{\delta} \psi_{(p)}$ identically satisfies the condition
\begin{align}
\bar{\delta}{\cal L} \equiv {J^\mu}_{:\mu} \label{sym0}
\end{align}
{\it off-shell} (i.e.\ without assuming that the equations of motion \eqref{m0} are satisfied) for some 4-vector $J^\mu$, then that transformation is a symmetry. In that case, it follows that
\begin{align}
\left( J^\mu-\sum_{p=1}^{P} \frac{\partial {\cal L}}{\partial (\psi_{(p),\mu})} \bar{\delta} \psi_{(p)} \right)_{:\mu}\equiv \sum_{p=1}^{P} \Lambda_{(\psi_{(p)})} \bar{\delta} \psi_{(p)}. \label{divc}
\end{align}
This is an identity valid off-shell. In particular, the right-hand side vanishes {\it on-shell} (i.e.\ when the equations of motion \eqref{m0} are satisfied), and Eq.\ \eqref{divc} becomes the conservation of the 4-current
\begin{align}
k^\mu=J^\mu-\sum_{p=1}^{P} \frac{\partial {\cal L}}{\partial (\psi_{(p),\mu})} \bar{\delta} \psi_{(p)}. \nonumber
\end{align}

Because the perturbed Lagrangian density ${\cal L}+ \bar{\delta}{\cal L}$ and the original Lagrangian density ${\cal L}$ differ only by a covariant divergence term when a symmetry provided by Eq.\ \eqref{sym0} exists, the action functional obtained from the perturbed Lagrangian density leads to the same equations of motion as that obtained from the original Lagrangian density. This is due to $\delta J^\mu$ (defined in the context of Hamilton's principle) vanishing at the 4-volume integration limits. The conserved current is associated with this invariance of the equations of motion. This is Noether's first theorem for active transformations of fields. Note however that \citet{Noether18} considered only Lagrangian densities that are strictly invariant under symmetry transformations (i.e.\ the particular case where $J^\mu=0$). The generalization to invariant Lagrangian densities {\it up to a divergence} is made in \citet{Bessel-Hagen21}, who states that he owes this generalization to an oral communication with E.\ Noether. \citet{Kosmann-Schwarzbach11} provides mathematical, physical, and historical perspectives on Noether's two theorems.

\subsection{Passive transformations}
Consider now an infinitesimal synchronous transformation of coordinates parametrized as $x^\mu \rightarrow \tilde{x}^{\mu}=x^\mu + \epsilon^\mu$, where $\epsilon^\mu$ is an infinitesimal 4-vector. The spatial components $\epsilon^i$ may be functions of the space-time coordinates, in other words the infinitesimal coordinate transformation of the spatial components is assumed to be local. In contrast, the time component $\epsilon^0$ must be at most an infinitesimally small constant due to the absolute nature of time intervals in Newtonian mechanics. In other words, only time shifting by a constant is admissible here.

By definition, a scalar field preserves its values at all space-time points under a coordinate transformation. Therefore, one may write $\tilde{\psi}_{(p)}(\tilde{x})=\psi_{(p)}(x)$. Here, $\tilde{x}$ and $x$ represent the same space-time point expressed in two different coordinate systems. In the coordinate system $\tilde{x}^\mu$, the functional form of the scalar is different from the functional form of the same scalar in the coordinate system $x^\mu$. This difference is symbolized by writing the scalar field as $\tilde{\psi}_{(p)}$ in the coordinate system $\tilde{x}^\mu$. To first order, one may therefore write
\begin{align}
\tilde{\psi}_{(p)}(\tilde{x})&=\tilde{\psi}_{(p)}(x) + \tilde{\psi}_{(p),\mu}(x) \epsilon^\mu, \\
&= \psi_{(p)}(x).
\end{align}
At a point $x$ of the original coordinate system, define the perturbations
\begin{align}
\tilde{\delta} \psi_{(p)} \equiv \tilde{\psi}_{(p)}(x) - \psi_{(p)}(x) = -\tilde{\psi}_{(p),\mu}(x) \epsilon^\mu = -\psi_{(p),\mu}(x) \epsilon^\mu. \label{dp}
\end{align}
The last equality stems from the fact that only first order perturbations are considered.

This shows that an infinitesimal coordinate transformation may be interpreted as a particular field transformation at a fixed point in space-time. This type of transformation is called a {\it passive} field transformation. The fields are transformed at fixed points within the original coordinate system as the result of an infinitesimal coordinate transformation.

The Lagrangian density depends on the scalar dynamical fields, but also on their first derivatives. The first derivatives of the scalar dynamical fields are covariant 4-vectors. Therefore, one must define how an infinitesimal coordinate transformation on a covariant 4-vector is written in terms of passive transformations. By definition, a covariant 4-vector $A_\mu$ transforms as
\begin{align}
A_\nu (x)=\frac{\partial \tilde{x}^\mu}{\partial x^\nu}\tilde{A}_{\mu} (\tilde{x}).
\end{align}
For the infinitesimal transformation $\tilde{x}^\mu=x^\mu+\epsilon^\mu$, one writes to first order
\begin{align}
A_\nu (x) &= (\delta^\mu_\nu+{\epsilon^\mu}_{,\nu})\tilde{A}_{\mu} (\tilde{x}), \\
&=\tilde{A}_{\nu} (\tilde{x}) + \tilde{A}_{\mu} (\tilde{x}) {\epsilon^\mu}_{,\nu}, \\
&=\tilde{A}_{\nu} (x) + \tilde{A}_{\nu,\mu} (x) \epsilon^\mu + \tilde{A}_{\mu} (x) {\epsilon^\mu}_{,\nu}, \\
&=\tilde{A}_{\nu} (x) + A_{\nu,\mu} (x) \epsilon^\mu + A_{\mu} (x) {\epsilon^\mu}_{,\nu}.
\end{align}
One may therefore define
\begin{align}
\tilde{\delta} A_\mu &\equiv \tilde{A}_{\mu} (x) - A_\mu (x), \\
&=-A_{\mu,\nu} (x) \epsilon^\nu - A_{\nu} (x) {\epsilon^\nu}_{,\mu}=-A_{\mu:\nu} (x) \epsilon^\nu - A_{\nu} (x) {\epsilon^\nu}_{:\mu}, \label{dcontra} \\
&=-(A_{\mu,\nu} (x)-A_{\nu,\mu} (x))\epsilon^\nu - (A_{\nu} (x) {\epsilon^\nu})_{,\mu}.
\end{align}
In the special case where $A_\mu$ is $\psi_{(p),\mu}$, the term $A_{\mu,\nu}-A_{\nu,\mu}$ vanishes because ordinary derivatives commute. This yields
\begin{align}
\tilde{\delta} (\psi_{(p),\mu}) = -(\psi_{(p),\nu} (x) {\epsilon^\nu})_{,\mu} = (\tilde{\delta} \psi_{(p)})_{,\mu}. \label{swp}
\end{align}
As was the case for active field transformations, the passive perturbation of the gradient of a scalar field is equal to the gradient of the passive perturbation.

An infinitesimal coordinate transformation interpreted as a passive transformation induces changes not only to the dynamical fields, but also to the metric tensor, its determinant, and to the external fields such as the gravitational potential. For passive transformations of the action functional, all perturbations occur within the original coordinate system $x^\mu$, and the integration limits are consequently kept fixed. Therefore,
\begin{align}
\tilde{\delta} {\cal S}=\tilde{\delta} \int d^4x \sqrt{g} {\cal L} = \int d^4x \, \tilde{\delta}(\sqrt{g} {\cal L}).
\end{align}
The variation $\tilde{\delta}(\sqrt{g} {\cal L})$ is written
\begin{align}
\tilde{\delta}(\sqrt{g} {\cal L}) = \sqrt{g} \sum_{p=1}^{P}\left(\frac{\partial {\cal L}}{\partial \psi_{(p)}} \tilde{\delta}\psi_{(p)} + \frac{\partial {\cal L}}{\partial (\psi_{(p),\mu})} \tilde{\delta} (\psi_{(p),\mu})\right) + \frac{\partial (\sqrt{g}{\cal L})}{\partial g^{\mu\nu}} \tilde{\delta} g^{\mu\nu} + \sqrt{g}\frac{\partial {\cal L}}{\partial \Phi} \tilde{\delta} \Phi,
\end{align}
where $\tilde{\delta} g^{\mu\nu}$ is given by Eq.\ \eqref{killing}, and $\tilde{\delta} \Phi=-\Phi_{,\mu}\epsilon^\mu$ as in Eq.\ \eqref{dp} since $\Phi$ is a scalar. It is convenient to define the symmetric tensor $t_{\mu\nu}$ as
\begin{align}
t_{\mu\nu} \equiv -\frac{2}{\sqrt{g}}\frac{\partial (\sqrt{g}{\cal L})}{\partial g^{\mu\nu}},
\end{align}
which is the covariant mass-momentum tensor $T_{\mu\nu}=\rho u_\mu u_\nu + h_{\mu\nu}p$ (where $p$ is pressure) plus a term that vanishes on-shell (see Appendix \ref{attt}). One may then rewrite the variation
\begin{align}
\tilde{\delta}(\sqrt{g} {\cal L}) = \sqrt{g} \left[ \sum_{p=1}^{P} \Lambda_{(\psi_{(p)})} \tilde{\delta}\psi_{(p)} - \frac{1}{2} t_{\mu\nu} \tilde{\delta} g^{\mu\nu} + \frac{\partial {\cal L}}{\partial \Phi} \tilde{\delta} \Phi + \left( \sum_{p=1}^{P}\frac{\partial {\cal L}}{\partial (\psi_{(p),\mu})} \tilde{\delta} \psi_{(p)} \right)_{:\mu}\right], \label{delta1}
\end{align}
from Eqs \eqref{motion} and \eqref{swp}.

In addition, the quantity $\tilde{\delta}(\sqrt{g} {\cal L})$ may always be rewritten as a divergence under a passive transformation. To see this, consider the perturbations of $\sqrt{g}$ and ${\cal L}$ separately. The former is expressed as
\begin{align}
\tilde{\delta}(\sqrt{g})=-\sqrt{g} {\epsilon^\mu}_{:\mu}=-(\sqrt{g} {\epsilon^\mu})_{,\mu}
\end{align}
(see Appendix \ref{pvs}), and the latter, being a scalar, as
\begin{align}
\tilde{\delta}{\cal L}=-{\cal L}_{,\mu} \epsilon^\mu,
\end{align}
similarly to Eq.\ \eqref{dp}. Therefore,
\begin{align}
\tilde{\delta}(\sqrt{g} {\cal L}) = -(\sqrt{g}{\cal L}\epsilon^\mu)_{,\mu}=-\sqrt{g}({\cal L}\epsilon^\mu)_{:\mu}. \label{delta2}
\end{align}
This expression is nothing but the statement that equations of motion are covariant: a coordinate transformation does not change the form of the equations of motion since the integrand defining the action functional changes by a total divergence under any infinitesimal coordinate transformation. In itself, this does not automatically translate into an actual symmetry. From Eqs \eqref{dp}, \eqref{delta1} and \eqref{delta2}, one writes
\begin{align}
\left( \left[ \sum_{p=1}^{P}\frac{\partial {\cal L}}{\partial (\psi_{(p),\mu})} \psi_{(p),\nu} -\delta_\nu^{\mu}{\cal L} \right] \epsilon^\nu  \right)_{:\mu} =\sum_{p=1}^{P} \Lambda_{(\psi_{(p)})} \tilde{\delta}\psi_{(p)} - \frac{1}{2} t_{\mu\nu} \tilde{\delta} g^{\mu\nu} + \frac{\partial {\cal L}}{\partial \Phi} \tilde{\delta} \Phi. \label{currentpass}
\end{align}
On-shell, Eq.\ \eqref{currentpass} represents the conservation of a 4-current
\begin{align}
j^\mu=\left( \sum_{p=1}^{P}\frac{\partial {\cal L}}{\partial (\psi_{(p),\mu})} \psi_{(p),\nu} -\delta_\nu^{\mu}{\cal L} \right) \epsilon^\nu
\end{align}
if the two symmetry conditions $\tilde{\delta} g^{\mu\nu}=0$ and $\tilde{\delta} \Phi=0$ are satisfied. These symmetry conditions on the metric tensor and external forcing are equivalent to
\begin{align}
\epsilon^{\mu:\nu}&=-\epsilon^{\nu:\mu}, \label{sym1}\\
\Phi_{,\mu} \epsilon^{\mu}&=0 \label{sym2}
\end{align}
(see Appendix \ref{kf} and Eq.\ \eqref{dp}).

Again, because the perturbed density $\sqrt{g}{\cal L}+\tilde{\delta}({\sqrt{g}\cal L})$ and the original density $\sqrt{g}{\cal L}$ differ by a divergence term only, the action functional obtained from the perturbed density leads to the same tensor equations of motion as that obtained from the original density\footnote{However, the new Lagrangian density ${\cal L}'={\cal L} - ({\cal L}\epsilon^\mu)_{:\mu}$ now also depends on the second derivatives of the scalar dynamical fields. In this case, the arbitrary variations defined in the context of Hamilton's principle $\delta\psi_{(p)}$ and $\delta(\psi_{(p),\mu})$ both vanish at the 4-volume integration limits, and the Euler-Lagrange equations take the form $$\frac{\partial {\cal L}'}{\partial \psi_{(p)}}-\left( \frac{\partial {\cal L}'}{\partial (\psi_{(p),\mu})} \right)_{:\mu}+\left( \frac{\partial {\cal L}'}{\partial (\psi_{(p),\mu:\nu})} \right)_{:\nu:\mu}=0.$$}. The conserved 4-current is associated with the symmetry conditions \eqref{sym1} and \eqref{sym2}. This is Noether's first theorem for infinitesimal coordinate transformations interpreted here as passive transformations on dynamical fields, the metric tensor, and the external forcing.

\subsection{Active and passive transformations combined}
Active and passive transformations may be combined from Eqs \eqref{divc} and \eqref{currentpass} to yield
\begin{align}
\left( J^\mu-\sum_{p=1}^{P} \frac{\partial {\cal L}}{\partial (\psi_{(p),\mu})} \left( \bar{\delta} \psi_{(p)} - \psi_{(p),\nu} \epsilon^\nu \right) - {\cal L} \epsilon^\mu\right)_{:\mu} = \sum_{p=1}^{P} \Lambda_{(\psi_{(p)})} (\bar{\delta}+\tilde{\delta})\psi_{(p)} - t_{\mu\nu} \epsilon^{\mu:\nu} - \frac{\partial {\cal L}}{\partial \Phi} \Phi_{,\mu} \epsilon^\mu.
\end{align}

In general, active and passive transformations are not special cases of the arbitrary variations defined in the context of Hamilton's least action principle: initial, final, and boundary field values are not necessarily kept fixed under active and passive transformations.

Recall that two Lagrangian densities that differ by a covariant divergence lead to the same equations of motion from Hamilton's least action principle. The transformation of $\sqrt{g}{\cal L}$ into
\begin{align}
\sqrt{g}{\cal L} + (\bar{\delta}+\tilde{\delta})(\sqrt{g}{\cal L}) = \sqrt{g}\left({\cal L}+\left(J^\mu-{\cal L}\epsilon^{\mu}\right)_{:\mu}\right),
\end{align}
following which the equations of motion are automatically unchanged, does not imply that the action functional ${\cal S}$ itself is necessarily unchanged. In the 4-volume integration that leads to a transformed action functional ${\cal S'}={\cal S}+(\bar{\delta}+\tilde{\delta}){\cal S}$, possible remaining boundary terms induced by the transformation mean that $(\bar{\delta}+\tilde{\delta}){\cal S}$ does not vanish in general. However, $(\bar{\delta}+\tilde{\delta}){\cal S}$ does not need to vanish for conserved currents to exist. The invariance of the equations of motion under the transformations and the realization of the symmetry conditions suffice.

Some often cited textbooks \citep[e.g.][section 13.7]{Goldstein02} present a simplified version of Noether's first theorem, in which conserved currents are associated with the strict invariance of the Lagrangian density. This may serve pedagogical purposes, but does not provide the most general form of the theorem. Since \citet{Bessel-Hagen21}, Noether's first theorem associates conserved currents with symmetries of the {\it equations of motion}, which is less restrictive than invariance of the Lagrangian density. Although \citet[][section 13.7]{Goldstein02} considered both active and passive transformations, their derivation does not account for contributions from the metric terms.

In sum, active transformations are symmetry transformations if a 4-vector $J^\mu$ exists such that $\bar{\delta} (\sqrt{g} {\cal L})=(\sqrt{g} J^\mu)_{,\mu}$ off-shell. For passive transformations, $\tilde{\delta} (\sqrt{g} {\cal L})$ can always be expressed as a total divergence because, in physics, acceptable equations of motion are covariant. Passive transformations are symmetry transformations if specific conditions on the external forcing and metric tensor are satisfied. The conservation laws associated with these active and passive transformations are called ``non-trivial'': they exist on-shell only, and they arise from internal or space-time symmetries of the equations of motion.

\subsection{Symmetries from active transformations of fields: mass and entropy conservation} \label{mec}
As a first example of symmetries from active field transformations that lead to conserved currents, consider the simple transformation of the Clebsch potential $\alpha$
\begin{align}
\bar{\delta} \alpha = \epsilon,
\end{align}
where $\epsilon$ is a small constant parameter. All other dynamical scalar fields are kept unchanged. This transformation leaves the vector field $v_\mu$ and the Lagrangian density unchanged ($\bar{\delta} v_\mu=0$ and $\bar{\delta}{\cal L}=0$) and is therefore a symmetry. From Eq.\ \eqref{sym0}, one may choose $J^\mu=0$. From Eqs \eqref{lag} and \eqref{divc},
\begin{align}
-\sum_{p=1}^{P} \frac{\partial {\cal L}}{\partial (\psi_{(p),\mu})} \bar{\delta} \psi_{(p)}&=\rho u^\mu \left( \bar{\delta} \alpha + \beta \bar{\delta} s + \sum_{r=1}^{N} \gamma_{(r)} \bar{\delta} \lambda_{(r)} \right), \label{currex}
\end{align}
i.e.\ the associated on-shell conserved 4-current is simply $\epsilon \rho u^\mu$. This should not come as a surprise since the variation of ${\cal L}$ with respect to $\alpha$ leads to the continuity equation.

As a second example, consider the more general transformations
\begin{align}
\bar{\delta} \rho &= 0, \\
\bar{\delta} \alpha &= \epsilon \left( {\cal F}-\sum_{r=1}^{N}\gamma_{(r)} \frac{\partial {\cal F}}{\partial \gamma_{(r)}} \right), \label{lgt1} \\
\bar{\delta} \beta &= -\epsilon \frac{\partial {\cal F}}{\partial s}, \\
\bar{\delta} s &= 0, \\
\bar{\delta} \gamma_{(r)} &= -\epsilon \frac{\partial {\cal F}}{\partial \lambda_{(r)}}, \\
\bar{\delta} \lambda_{(r)} &= \epsilon \frac{\partial {\cal F}}{\partial \gamma_{(r)}}, \label{lgt2}
\end{align}
where $\epsilon$ is again a small constant parameter and ${\cal F}={\cal F}(s,\gamma_{(1)},\lambda_{(1)},...,\gamma_{(N)},\lambda_{(N)})$ is any differentiable function that depends locally on $s$, $\gamma_{(r)}$ and $\lambda_{(r)}$, but not on their space-time derivatives. It may be verified that these transformations also leave the vector field $v_\mu$ and the Lagrangian density unchanged ($\bar{\delta} v_\mu=0$ and $\bar{\delta}{\cal L}=0$), and one may again choose $J^\mu=0$ from Eq.\ \eqref{sym0}. From Eq.\ \eqref{currex}, the associated on-shell conserved 4-current is $\epsilon \rho u^\mu {\cal F}$. In particular, the choice ${\cal F}=s$ leads to the conservation of entropy, and ${\cal F}=1$ to the conservation of mass (see first example).

These specific symmetries, although instructive, are not particularly interesting as they lead to conservation laws already known directly from the equations of motion.

\subsection{Space-time symmetries: momentum and energy conservation}
Continuous space-time transformations are not necessarily symmetries in the sense of Noether. For continuous space-time transformations to be symmetry transformations, specific and concurrent conditions on the metric tensor and external forcing must be satisfied. In this case, these conditions are given by Eqs \eqref{sym1}--\eqref{sym2}.

Conserved 4-currents arising from purely space-time symmetries are obtained from Eq.\ \eqref{currentpass} for passive transformations. From the Lagrangian density \eqref{lag} and the definitions \eqref{vvv} and \eqref{vtou}, one obtains
\begin{align}
\sum_{p=1}^{P}\frac{\partial {\cal L}}{\partial (\psi_{(p),\mu})} \psi_{(p),\nu}=-\rho u^{\mu}v_{\nu}.
\end{align}
Define an energy density 4-vector similar to eq.\ (45) in \citet{Charron14a} as
\begin{align}
E^{\mu} \equiv \rho u^{\mu}\left( \frac{1}{2} h^{\alpha\beta}u_\alpha u_\beta +I + \Phi' \right) + u_\nu h^{\mu\nu} p
\end{align}
($\Phi' \equiv \Phi+1$ and $\Phi$ obviously represent the same gravitational field). The 4-current is written, after some algebra and from eq.\ (65) in \citet{Zadra15}, as
\begin{align}
j^\mu &\equiv \sum_{p=1}^{P}\frac{\partial {\cal L}}{\partial (\psi_{(p),\mu})} \psi_{(p),\nu} \epsilon^\nu - {\cal L}\epsilon^\mu, \\
&= -\rho u^\mu v_\nu \epsilon^\nu -p\epsilon^\mu- \rho \Lambda_{(\rho)}\epsilon^\mu, \\
&= \epsilon^0 E^\mu - \epsilon^\nu T^\mu_\nu + \rho \Lambda_{(\rho)}(\epsilon^0 u^\mu-\epsilon^\mu).
\end{align}
On-shell, Eq.\ \eqref{currentpass} becomes
\begin{align}
\left( \epsilon^0 E^\mu - \epsilon^\nu T^\mu_\nu \right)_{:\mu}= -T_{\mu\nu}\epsilon^{\mu:\nu}+\rho\Phi_{,\mu}\epsilon^\mu. \label{emc}
\end{align}
If the symmetry conditions \eqref{sym1} and \eqref{sym2} are satisfied, the right-hand side vanishes (recall that $T_{\mu\nu}=T_{\nu\mu}$ would then be contracted with an antisymmetric tensor $\epsilon^{\mu:\nu}=-\epsilon^{\nu:\mu}$) and the 4-current $\epsilon^0 E^\mu - \epsilon^\nu T^\mu_\nu$ becomes covariant divergent-free.

Suppose that one chooses $\epsilon^0=0$. It has been shown by \citet{Charron14b} that the symmetry conditions \eqref{sym1} and \eqref{sym2} are satisfied when having $g_{\mu\nu,i}=0$, $\Phi_{,i}=0$, $\epsilon^i=\text{constant}$ ($i=1 \text{ or } 2 \text{ or } 3$), and $\epsilon^k=0$ (for any $k \ne i$) concurrently in at least one coordinate system. In that case, Eq.\ \eqref{emc} represents the conservation of momentum (linear, angular, or other) in any admissible coordinate system.

Similarly, if one chooses $\epsilon^0 \ne 0$, the symmetry conditions \eqref{sym1} and \eqref{sym2} are satisfied when having $g_{\mu\nu,0}=0$, $\Phi_{,0}=0$, and $\epsilon^k=0$ concurrently in at least one coordinate system \citep[see][]{Charron15a}. In that case, Eq.\ \eqref{emc} represents the conservation of energy in any admissible coordinate system.

In sum, in a given coordinate system conservation equations for momentum or energy exist if both the gravitational potential and the metric tensor exhibit the appropriate symmetries.

\section{Ertel's theorem and trivially conserved 4-currents} \label{secertel}
In the preceding section, mass and entropy conservation has been associated with an internal symmetry of the equations of motion, while momentum and energy conservation has been associated with space-time symmetries of the metric tensor and external forcing. Another fundamental conservation law in geophysical fluid dynamics is that of Ertel's potential vorticity. Is there a symmetry of the equations of motion associated with Ertel's potential vorticity conservation? A vast body of scientific literature associates Ertel's potential vorticity conservation with the particle-relabeling symmetry, initially called exchange symmetry \citep[][and others]{Newcomb67,Bretherton70,Ripa81,Salmon82,Salmon88,Salmon98,Salmon13,Muller95,Padhye96}. It is demonstrated here that this association is unfounded. Studying this question in arbitrary coordinates provides insights that illustrate some advantages of the 4-dimensional tensor form over more traditional approaches.

\subsection{On the triviality of Ertel's potential vorticity conservation}
First, consider a generic antisymmetric tensor $F^{\mu\nu}=-F^{\nu\mu}$. Define a 4-vector $c^\mu$ as the covariant divergence of this antisymmetric tensor:
\begin{align}
c^\mu\equiv {F^{\mu\nu}}_{:\nu}. \nonumber
\end{align}
By virtue of Eq.\ \eqref{i3}, the covariant divergence of $c^\mu$ identically vanishes:
\begin{align}
{c^\mu}_{:\mu}={F^{\mu\nu}}_{:\nu:\mu}=0. \label{secondkind}
\end{align}
Following the terminology of \citet[][p.\ 264-265]{Olver93}, Eq.\ \eqref{secondkind} is a trivial conservation law of the second kind. Such trivial conservation laws are characterized by currents solely written in terms of the divergence of an antisymmetric tensor. They are mathematical identities obtained off-shell --- the equations of motion are not required to establish such trivial conservation laws --- and therefore they exist independently of symmetries of the equations of motion.

Antisymmetric quantities such as $F^{\mu\nu}$ are sometimes referred to as ``superpotentials''. Conserved currents may always be defined up to the divergence of a superpotential.

\subsubsection{Triviality of Ertel's potential vorticity conservation using Clebsch potentials}
In tensor notation, Ertel's potential vorticity $q$ is defined as
\begin{align}
q&=\frac{\varepsilon^{0\mu\nu\sigma}}{\sqrt{g}\rho} u_{\sigma:\nu} s_{,\mu} = \frac{\varepsilon^{0\mu\nu\sigma}}{\sqrt{g}\rho} v_{\sigma:\nu} s_{,\mu}\nonumber \\
&=\sum_{r=1}^{N}\frac{\varepsilon^{0\mu\nu\sigma}}{\sqrt{g}\rho}s_{,\mu} \gamma_{(r),\nu} \lambda_{(r),\sigma}.
\end{align}
Define
\begin{align}
q_{(r)}=\frac{\varepsilon^{0\mu\nu\sigma}}{\sqrt{g}\rho}s_{,\mu} \gamma_{(r),\nu} \lambda_{(r),\sigma}
\end{align}
(no sum over $r$), a 4-vector $A^\mu_{(r)}$ as
\begin{align}
A^\mu_{(r)} = \frac{\varepsilon^{0\mu\nu\sigma}}{\sqrt{g}}s \gamma_{(r),\nu} \lambda_{(r),\sigma}
\end{align}
(again, no sum over $r$), and the antisymmetric tensor $F^{\beta\mu}_{(r)}$ as
\begin{align}
F^{\beta\mu}_{(r)}=u^\beta A^\mu_{(r)}-u^\mu A^\beta_{(r)}. \label{spf}
\end{align}
The 4-current $\rho u^\beta q_{(r)}$ may be written as
\begin{align}
\rho u^\beta q_{(r)} &= \frac{\varepsilon^{0\mu\nu\sigma}}{\sqrt{g}}u^\beta s_{,\mu} \gamma_{(r),\nu} \lambda_{(r),\sigma} \nonumber \\
&= (u^\beta A^\mu_{(r)})_{:\mu}-{u^\beta}_{:\mu} A^\mu_{(r)} .
\end{align}
By virtue of Eqs \eqref{i1} and \eqref{i2},
\begin{align}
{u^\beta}_{:\mu} A^\mu_{(r)}&=\frac{1}{\sqrt{g}}s \gamma_{(r),\nu} \lambda_{(r),\sigma} \left( \varepsilon^{0\beta\nu\sigma} {u^\mu}_{:\mu} - \varepsilon^{0\beta\mu\sigma} {u^\nu}_{:\mu} - \varepsilon^{0\beta\nu\mu} {u^\sigma}_{:\mu}\right), \nonumber \\
&= (u^\mu A^\beta_{(r)})_{:\mu}-\frac{\varepsilon^{0\beta\nu\sigma}}{\sqrt{g}} \left( \dot{s} \gamma_{(r),\nu} \lambda_{(r),\sigma} + s \dot{\gamma}_{(r),\nu} \lambda_{(r),\sigma} + s \gamma_{(r),\nu} \dot{\lambda}_{(r),\sigma} \right).
\end{align}
One may therefore define a 4-current $c^\mu_{(r)}$ as
\begin{align}
c^\mu_{(r)} &\equiv \rho u^\mu q_{(r)} -\frac{\varepsilon^{0\mu\nu\sigma}}{\sqrt{g}} \left( \dot{s} \gamma_{(r),\nu} \lambda_{(r),\sigma} + s \dot{\gamma}_{(r),\nu} \lambda_{(r),\sigma} + s \gamma_{(r),\nu} \dot{\lambda}_{(r),\sigma} \right), \nonumber \\
&= \rho u^\mu q_{(r)} +\frac{\varepsilon^{0\mu\nu\sigma}}{\sqrt{g}} \Bigg( \rho^{-1}\gamma_{(r),\nu} \lambda_{(r),\sigma} \Lambda_{(\beta)} + s \gamma_{(r),\nu} \left( \rho^{-1} \Lambda_{(\gamma_{(r)})} \right)_{,\sigma} \nonumber \\
&\qquad\qquad\qquad\qquad\qquad - s \lambda_{(r),\sigma} \left( \rho^{-1} \Lambda_{(\lambda_{(r)})} - \rho^{-1} \gamma_{(r)}\Lambda_{(\alpha)} \right)_{,\nu} \Bigg), \nonumber \\
&= F^{\mu\nu}_{(r):\nu}. \label{pvcurrent}
\end{align}
This 4-current is therefore identically divergent-free ($c^\mu_{(r):\mu}=0$) by virtue of Eq.\ \eqref{i3}. On-shell, it becomes $\rho u^\mu q_{(r)}$. In the case of the current described by Eq.\ \eqref{pvcurrent}, the ``charge density'' $\sqrt{g}c^0_{(r)}$ is $\sqrt{g}\rho q_{(r)}$ off-shell.

Because the conservation equation
\begin{align*}
{c^\mu}_{:\mu}=\left( \sum_{r=1}^N c^\mu_{(r)} \right)_{:\mu}=0
\end{align*}
is a trivial conservation law of the second kind, Ertel's potential vorticity conservation cannot be associated with a particular symmetry of the equations of motion. Noether's first theorem does not apply in this case.

Noether's second theorem provides differential identities, if they exist, among the $\Lambda_{(\psi_{(p)})}$'s for under-determined dynamical systems \citep[][p.\ 342 and following]{Olver93}. Local gauge symmetries are related to Noether's second theorem. As will be shown in sub-section \ref{ngs} using Dirac's theory for constrained Hamiltonian systems, geophysical fluid dynamics (unapproximated and geometrically approximated) has no local gauge symmetries: the $\Lambda_{(\psi_{(p)})}$'s are independent from each other and the dynamical system is not under-determined. Therefore, the conservation of Ertel's potential vorticity is not obtained from Noether's second theorem either.

\subsubsection{Triviality of Ertel's potential vorticity conservation without Clebsch potentials}
The trivial nature of Ertel's potential vorticity conservation may also be demonstrated using more traditional dynamical fields instead of Clebsch potentials. If one chooses $(\rho,s,u^1,u^2,u^3)$ as dynamical fields, it is shown that
\begin{align}
\rho u^\alpha q + \frac{\omega^\alpha \Lambda_{(\beta)}}{\rho} - \frac{\varepsilon^{0\alpha \mu \sigma}}{\sqrt{g}} \frac{(\Lambda_\mu-u_\mu \Lambda^0)s_{,\sigma}}{\rho} = {G^{\alpha\mu}}_{:\mu}, \label{curqobs}
\end{align}
where
\begin{align}
\Lambda_{(\beta)} &= -\rho u^\mu s_{,\mu}, \label{ls} \\
\Lambda^{\mu} &= {T^{\mu\nu}}_{:\nu} + \rho h^{\mu\nu}\Phi_{,\nu}, \\
\omega^\alpha  &= \frac{\varepsilon^{0\alpha \mu \sigma}}{\sqrt{g}} u_{\sigma:\mu}, \label{omeg} \\
G^{\alpha\mu}  &= u^\alpha B^\mu - u^\mu B^\alpha + \frac{\varepsilon^{0\alpha \mu \sigma}}{\sqrt{g}} \left[ \left(\frac{1}{2}g_{\nu\beta}u^\nu u^\beta-\Phi-I-\frac{p}{\rho} \right)s_{,\sigma} + u_\sigma \frac{\Lambda_{(\beta)}}{\rho} \right], \label{gbm} \\
B^\mu         &= \frac{\varepsilon^{0\mu \nu \sigma}}{\sqrt{g}} u_\nu s_{,\sigma} \label{bbb}
\end{align}
(see Appendix \ref{trivobs}). The fact that $G^{\alpha\mu}$ is antisymmetric ensures that the left-hand side of Eq.\ \eqref{curqobs} is a trivially conserved 4-current of the second kind. The ``charge density'' associated with that 4-current is $\sqrt{g}\rho q$ off-shell. The terms $\Lambda_{\mu}=g_{\mu\nu}\Lambda^{\nu}$, $\Lambda^0$ and $\Lambda_{(\beta)}$ all vanish on-shell. The trivial nature of a conservation law obviously does not depend on the choice of dynamical fields. As for Eq.\ \eqref{pvcurrent}, the mathematical identity \eqref{curqobs} is demonstrated without assuming that the equations of motion are satisfied and therefore without assuming any symmetry of the equations of motion.

\citet{Rosenhaus16} also analyzed the trivial nature of Ertel's theorem in the context of approximated flows.

\subsection{Revisiting the derivation by \citet{Padhye96}}
In a particle-like Lagrangian formulation, Ertel's potential vorticity conservation has in the past been associated with the particle-relabeling symmetry by \citet{Newcomb67,Bretherton70,Ripa81,Salmon82,Salmon88,Salmon98,Salmon13,Muller95,Padhye96} and others. In particular, \citet{Padhye96} refer to Noether's two theorems to justify the association between particle-relabeling and Ertel's theorem. Is there a contradiction between, on the one hand, the fact that the conservation of Ertel's potential vorticity is a trivial law dissociated from any symmetry of the equations of motion in a field formulation with arbitrary coordinates; and on the other hand, the association between Ertel's potential vorticity conservation and a relabeling symmetry in the particle-like formulation? In other words, can a trivial conservation law in one given formulation become non-trivial in another formulation? It has previously been mentioned that the symmetry associated with Ertel's theorem is invisible in an Eulerian formulation, but exists in a particle-following formulation \citep[see for example][]{Shepherd03}. This question is revisited below.

\subsubsection{``Relabeling'' in arbitrary coordinates} 
In the following, the relabeling transformation proposed by \citet{Padhye96} is applied in the context of a field theory explicitly obeying the principle of Newtonian relativity, i.e.\ manifestly covariant for any admissible coordinate transformations in Newtonian mechanics. This manifestly covariant approach helps resolve the potential contradiction mentioned in the preceding paragraph.

The 4-volume integral over a fixed domain ${\cal D}$ of any scalar function $f$ is written
\begin{align}
{\cal A}=\int_{\cal D} d^4x \sqrt{g} f.
\end{align}
A coordinate transformation $\tilde{x}^\mu=x^\mu+\epsilon^\mu$, interpreted as a passive transformation of the metric terms and of $f$ results in a variation of ${\cal A}$:
\begin{align}
\tilde{\delta} {\cal A}&=-\int_{\cal D} d^4x \left( \sqrt{g} f \epsilon^\mu \right)_{,\mu}, \nonumber \\
&=-\oint_{\partial {\cal D}} dS_\mu f \epsilon^\mu,
\end{align}
from the divergence theorem in four dimensions. If the coordinate transformation is chosen such that $\epsilon^\mu=0$ at the integration limits $\partial {\cal D}$, then $\tilde{\delta} {\cal A}=0$.

Consider the following choices for $f$ and $\epsilon^\mu$:
\begin{align}
f&=-{\cal L}=\rho \left( \frac{1}{2}h^{\mu\nu}v_\mu v_{\nu} + g^{0\mu}v_\mu +I+\Phi \right), \\
\epsilon^\mu&=\epsilon \frac{\varepsilon^{0\mu\nu\sigma}}{\sqrt{g}\rho} s_{,\nu} C_{(1),\sigma}, \label{passiveq}
\end{align}
where $C_{(1)}=C_{(1)}(\lambda_{(1)})$ is an arbitrary local function of $\lambda_{(1)}$, and $\epsilon$ is an infinitesimally small constant. Recall from \citet{Zadra15} that the $\lambda_{(r)}$'s originate from the labels in the particle-like formulation. The function $C_{(1)}$, although arbitrary inside the domain ${\cal D}$, is chosen to have vanishing derivative $C_{(1),\sigma}$ at its boundary $\partial {\cal D}$. The $\epsilon^\mu$ of Eq.\ \eqref{passiveq} corresponds to the tensor form of eq.\ (35) in \citet{Padhye96}. These choices lead to
\begin{align}
\tilde{\delta}(\sqrt{g}\rho)&=-\sqrt{g}(\rho\epsilon^\mu)_{:\mu}=0, \\
\tilde{\delta}s&=0.
\end{align}
The variation of the action functional may be decomposed into two vanishing scalar terms
\begin{align}
\tilde{\delta} {\cal S}&=\tilde{\delta} {\cal S}_K + \tilde{\delta} {\cal S}_P,
\end{align}
where
\begin{align}
\tilde{\delta} {\cal S}_K &= \int_{\cal D} d^4x \sqrt{g} \rho \; \tilde{\delta}\left( \frac{1}{2}h^{\mu\nu}v_\mu v_{\nu} + g^{0\mu}v_\mu \right) =0,\\
\tilde{\delta} {\cal S}_P &= \int_{\cal D} d^4x \sqrt{g} \rho \; \tilde{\delta}\left( I+\Phi \right) =0.
\end{align}

It can be shown that the term $\tilde{\delta} {\cal S}_K$ may be rewritten as the sum of two terms: $\tilde{\delta} {\cal S}_K=\tilde{\delta} {\cal S}_{KD}+\tilde{\delta} {\cal S}_{KM}=0$, where
\begin{align}
\tilde{\delta} {\cal S}_{KD}&=\int_{\cal D} d^4x \sqrt{g} \rho u^\mu \tilde{\delta} v_\mu, \\
\tilde{\delta} {\cal S}_{KM}&=\int_{\cal D} d^4x \sqrt{g} \rho u^\mu v_\nu {\epsilon^\nu}_{:\mu}.
\end{align}
The term $\tilde{\delta} {\cal S}_{KD}$ comes from the passive variation of dynamical fields, while $\tilde{\delta} {\cal S}_{KM}$ comes from the passive variation of the metric tensor. It is important to note that $\tilde{\delta} {\cal S}_{KD}$ and $\tilde{\delta} {\cal S}_{KM}$ are non-vanishing terms off-shell.

Consider the integrand of $\tilde{\delta} {\cal S}_{KD}$:
\begin{align}
\sqrt{g}\rho u^\mu \tilde{\delta} v_\mu &=\sqrt{g}\rho u^\mu \left( (\tilde{\delta}\alpha)_{,\mu} +s_{,\mu}\tilde{\delta}\beta + \sum_{r=1}^{N} \left[ \lambda_{(r),\mu}\tilde{\delta}\gamma_{(r)} + \gamma_{(r)} (\tilde{\delta}\lambda_{(r)})_{,\mu} \right]\right), \nonumber \\
&=\sqrt{g}\epsilon^\mu\left( \Lambda_{(\alpha)}\alpha_{,\mu} +\Lambda_{(\beta)}\beta_{,\mu} +\sum_{r=1}^{N} \Lambda_{(\lambda_{(r)})}\lambda_{(r),\mu} + \sum_{r=2}^{N} \Lambda_{(\gamma_{(r)})}\gamma_{(r),\mu} \right) \nonumber \\
&\quad -\sqrt{g}\rho u^\mu \epsilon^\nu \lambda_{(1),\mu} \gamma_{(1),\nu} + \text{divergent terms.}
\end{align}
One may rearrange the preceding equation by noting that
\begin{align}
\epsilon^\mu \Lambda_{(\psi_{(p)})}\psi_{(p),\mu} &= \epsilon \frac{\varepsilon^{0\mu\nu\sigma}}{\sqrt{g}\rho} s_{,\nu} C_{(1),\sigma} \Lambda_{(\psi_{(p)})}\psi_{(p),\mu}, \nonumber \\
&=C_{(1)}\left( \epsilon \frac{\varepsilon^{0\mu\nu\sigma}}{\sqrt{g}\rho} s_{,\nu} \Lambda_{(\psi_{(p)})}\psi_{(p),\sigma}  \right)_{:\mu} + \text{covariant divergent terms,} \\
\rho u^\mu\epsilon^\nu \lambda_{(1),\mu}\gamma_{(1),\nu} &= \epsilon u^\mu \frac{\varepsilon^{0\nu\tau\sigma}}{\sqrt{g}} s_{,\tau} C_{(1),\sigma} \lambda_{(1),\mu}\gamma_{(1),\nu}, \nonumber \\
&= \epsilon u^\mu \frac{\varepsilon^{0\nu\tau\sigma}}{\sqrt{g}} s_{,\tau} \frac{dC_{(1)}}{d\lambda_{(1)}} \lambda_{(1),\sigma} \lambda_{(1),\mu}\gamma_{(1),\nu}, \nonumber \\
&= \epsilon u^\mu \frac{\varepsilon^{0\nu\tau\sigma}}{\sqrt{g}} s_{,\tau} C_{(1),\mu} \lambda_{(1),\sigma}\gamma_{(1),\nu}, \nonumber \\
&= -\epsilon \rho u^\mu q_{(1)} C_{(1),\mu}, \nonumber \\
&= C_{(1)}\left( \epsilon \rho u^\mu q_{(1)}  \right)_{:\mu} + \text{covariant divergent terms.}
\end{align}
One may therefore write $\tilde{\delta} {\cal S}_{KD}$ as
\begin{align}
\tilde{\delta} {\cal S}_{KD}= -\int_{\cal D} d^4x \sqrt{g} C_{(1)} {Q^\mu_{(1):\mu}}, \label{certel}
\end{align}
where
\begin{align}
Q^\mu_{(1)}&=\epsilon \rho u^\mu q_{(1)} - \epsilon \frac{\varepsilon^{0\mu\nu\sigma}}{\sqrt{g}\rho} s_{,\nu} \times \nonumber \\
&\left( \Lambda_{(\alpha)}\alpha_{,\sigma} + \Lambda_{(\beta)}\beta_{,\sigma} - \Lambda_{(\gamma_{(1)})}\gamma_{(1),\sigma}+\sum_{r=1}^N \left[ \Lambda_{(\gamma_{(r)})}\gamma_{(r),\sigma} +\Lambda_{(\lambda_{(r)})}\lambda_{(r),\sigma} \right] \right).
\end{align}
It is shown explicitly in Appendix \ref{secondt} that $\tilde{\delta} {\cal S}_{KM}$ cancels $\tilde{\delta} {\cal S}_{KD}$ identically, as it must. The same procedure may be followed for $Q^\mu_{(t)}$ ($t=2,...,N$).

The variation of the metric tensor, which leads to the term $\tilde{\delta} {\cal S}_{KM}$, must not be overlooked. Its omission would lead to the erroneous conclusion that Ertel's potential vorticity conservation is associated with the transformation \eqref{passiveq}: if $C_{(r)}$ is arbitrary and if $\tilde{\delta} {\cal S}_{KD}$ vanishes, then $\sum_{r=1}^N{Q^\mu_{(r):\mu}}$ vanishes from the du Bois-Reymond lemma. However, when one considers the complete variation of the action functional under Eq.\ \eqref{passiveq} (i.e.\ dynamical fields {\it and} metric tensor), one ends up with a trivial identity $0=0$ that does not reveal anything about conservation laws.

Note also that the transformation given by Eq.\ \eqref{passiveq} is not a symmetry transformation as defined in the context of Noether's first theorem. Although the term $(\partial {\cal L}/\partial \Phi)\tilde{\delta} \Phi$ may be rewritten, from $(\rho\epsilon^\mu)_{:\mu}=0$, as
\begin{align}
\frac{\partial {\cal L}}{\partial \Phi} \tilde{\delta} \Phi=(\rho \epsilon^\mu \Phi)_{:\mu},
\end{align}
allowing one to absorb $-\rho \epsilon^\mu \Phi$ into the definition of a 4-current, the term $\tilde{\delta} g^{\mu\nu}=\epsilon^{\mu:\nu}+\epsilon^{\nu:\mu}$ however does not vanish. In arbitrary coordinates, the passive variation of the metric tensor prevents the existence of a ``relabeling'' symmetry.

\subsubsection{Triviality of Ertel's potential vorticity conservation in label representation}
Ertel's potential vorticity conservation may also be studied in label representation within a particle-like formulation. In such formulation, the fluid particles may be labeled with any admissible curvilinear coordinate system ${\mathbf a}=(a^1,a^2,a^3)$ at a given time $\tau=0$. A particle mass is provided by $da^1da^2da^3\sqrt{{\cal G}} \rho_0 =d^3a\sqrt{{\cal G}} \rho_0$, where ${\cal G}={\cal G}({\mathbf a})$ is the determinant of the covariant metric tensor in coordinates $(\tau,a^1,a^2,a^3)$ at $\tau=0$, and $\rho_0$ is the particle's density at $\tau=0$. At a given point in space-time, the mass of a fluid element $da^1da^2da^3\sqrt{{\cal G}} \rho_0$ is equal to $dx^1dx^2dx^3\sqrt{g} \rho$ in arbitrary coordinates $(x^0,x^1,x^2,x^3)$. This ensures the {\it a priori} conservation of total mass. In addition, the material conservation of specific entropy is assumed {\it a priori}, therefore $s\equiv s({\mathbf a})$ and $\partial s/\partial \tau=0$.

The expression for Ertel's potential vorticity $q$ is translated from arbitrary coordinates to label representation as follows:
\begin{align}
q&=\frac{\varepsilon^{0ijk}}{\sqrt{g}\rho}s_{,k}u_{j:i}, \nonumber \\
&=\frac{\varepsilon^{0ijk}}{\sqrt{g}\rho}s_{,k}u_{j,i}, \nonumber \\
&=\frac{\varepsilon^{0ijk}}{\sqrt{g}\rho}\frac{\partial s}{\partial a^n}\frac{\partial a^n}{\partial x^k}\frac{\partial u_j}{\partial a^l}\frac{\partial a^l}{\partial x^i}, \nonumber \\
&=\frac{\varepsilon^{0lmn}}{\sqrt{{\cal G}}\rho_0}\frac{\partial s}{\partial a^n}\frac{\partial x^j}{\partial a^m}\frac{\partial u_j}{\partial a^l},
\end{align}
where the relation
\begin{align}
\frac{\varepsilon^{0ijk}}{\sqrt{g}\rho} \frac{\partial a^l}{\partial x^i} \frac{\partial a^n}{\partial x^k} = \frac{\varepsilon^{0lmn}}{\sqrt{{\cal G}}\rho_0} \frac{\partial x^j}{\partial a^m}
\end{align}
representing the {\it a priori} constraint on mass conservation has been used.

Consider now the antisymmetric tensor $A^{\mu\nu}$ defined as
\begin{align}
A^{\mu\nu}\equiv \frac{\varepsilon^{\mu\nu\alpha\beta}}{\sqrt{{\cal G}}} \frac{\partial s}{\partial a^\beta}\left( \frac{\partial x^\sigma}{\partial a^\alpha} u_\sigma - \delta^0_\alpha \left[ \frac{1}{2}u^\sigma u_\sigma+I+\Phi+\frac{p}{\rho}\right] \right) \nonumber
\end{align}
with explicit components
\begin{align}
A^{00}&=0, \\
A^{0l}&=\frac{\varepsilon^{0lmn}}{\sqrt{{\cal G}}}\frac{\partial s}{\partial a^n}\frac{\partial x^j}{\partial a^m} u_j=\frac{\varepsilon^{0lmn}}{\sqrt{{\cal G}}}\frac{\partial s}{\partial a^n}\frac{\partial x^\mu}{\partial a^m} u_\mu=-A^{l0}, \\
A^{lm}&=\frac{\varepsilon^{0lmn}}{\sqrt{{\cal G}}}\frac{\partial s}{\partial a^n}\left( \frac{1}{2}g_{\mu\nu}u^\mu u^\nu-I-\Phi-\frac{p}{\rho}  \right),
\end{align}
where $a^0=x^0=\tau$, $u^0=1$, $u^i=\partial x^i({\mathbf a},\tau) /\partial \tau$, and $g_{\mu\nu}=g_{\mu\nu}(x^i({\mathbf a},\tau),\tau)$.

Define the 4-vector $F^\mu$ as
\begin{align}
F^\mu \equiv \frac{1}{\sqrt{{\cal G}}} \frac{\partial}{\partial a^\nu} \left(\sqrt{{\cal G}}A^{\mu\nu}\right). \label{defflux}
\end{align}
The component $F^0$ is
\begin{align}
F^0=\frac{1}{\sqrt{{\cal G}}}\frac{\partial}{\partial a^\nu} \left(\sqrt{{\cal G}}A^{0\nu}\right)=\frac{1}{\sqrt{{\cal G}}}\frac{\partial}{\partial a^l} \left(\sqrt{{\cal G}}A^{0l}\right)=\rho_0 q,
\end{align}
and the components $F^l$ are
\begin{align}
\sqrt{{\cal G}}F^l&=\frac{\partial}{\partial a^\nu} \left(\sqrt{{\cal G}}A^{l\nu}\right)=-\frac{\partial}{\partial \tau} \left(\sqrt{{\cal G}}A^{0l}\right) + \frac{\partial}{\partial a^m} \left(\sqrt{{\cal G}}A^{lm}\right), \nonumber \\
&=-\varepsilon^{0lmn}\frac{\partial s}{\partial a^n}\left[ \frac{\partial u^\mu}{\partial a^m} u_\mu + \frac{\partial x^j}{\partial a^m} \frac{\partial u_j}{\partial \tau}-\frac{\partial}{\partial a^m}\left( \frac{1}{2}g_{\mu\nu}u^\mu u^\nu-I-\Phi-\frac{p}{\rho}  \right)\right], \nonumber \\
&=-\varepsilon^{0lmn}\frac{\partial s}{\partial a^n}\left[ \frac{\partial x^j}{\partial a^m} \frac{\partial u_j}{\partial \tau} -\frac{1}{2}u^\mu u^\nu \frac{\partial g_{\mu\nu}}{\partial a^m} + \frac{\partial \Phi}{\partial a^m} + \frac{1}{\rho}\frac{\partial p}{\partial a^m} \right], \nonumber \\
&=-\varepsilon^{0lmn}\frac{\partial s}{\partial a^n} \frac{\partial x^j}{\partial a^m} \left[ \frac{\partial u_j}{\partial \tau} -\frac{1}{2}u^\mu u^\nu \frac{\partial g_{\mu\nu}}{\partial x^j} + \frac{\partial \Phi}{\partial x^j} + \frac{1}{\rho}\frac{\partial p}{\partial x^j} \right].
\end{align}
Note that the quantity in brackets, if equated to zero, represents the equations of motion for the covariant momentum in arbitrary coordinates. One may write
\begin{align}
\Lambda_{(x^j)} \equiv -\rho_0 \left( \frac{\partial u_j}{\partial \tau} -\frac{1}{2}u^\mu u^\nu \frac{\partial g_{\mu\nu}}{\partial x^j} + \frac{\partial \Phi}{\partial x^j} + \frac{1}{\rho}\frac{\partial p}{\partial x^j}  \right) \label{motionlabel}
\end{align}
with $\partial u_j / \partial \tau = d u_j / dt = u^\mu \partial u_j / \partial x^\mu$ \citep[see][]{Zadra15}, and the components $F^l$ thus become
\begin{align}
F^l=\frac{\varepsilon^{0lmn}}{\sqrt{{\cal G}}\rho_0}\frac{\partial s}{\partial a^n} \frac{\partial x^j}{\partial a^m} \Lambda_{(x^j)}
\end{align}
with $\Lambda_{(x^j)}$ vanishing on-shell. The quantity $\partial ({\sqrt{{\cal G}}}F^\mu) / \partial a^\mu$ therefore takes the form
\begin{align}
\frac{\partial}{\partial a^\mu} \left(\sqrt{{\cal G}}F^\mu\right)=\frac{\partial }{\partial \tau}\left(\sqrt{{\cal G}}\rho_0 q\right)+ \frac{\partial}{\partial a^l} \left( \frac{\varepsilon^{0lmn}}{\rho_0}\frac{\partial s}{\partial a^n} \frac{\partial x^j}{\partial a^m} \Lambda_{(x^j)}  \right) = \frac{\partial^2}{\partial a^\mu \partial a^\nu} \left(\sqrt{{\cal G}}A^{\mu\nu}\right)=0. \label{defolver}
\end{align}
It vanishes due to the antisymmetry of $A^{\mu\nu}$ and the commutativity of ordinary partial derivatives. This conservation equation in label representation is a mathematical identity: it has been obtained off-shell, i.e.\ independently of the equations of motion, and therefore independently of any assumed symmetry of these equations. As discussed earlier, \citet{Olver93} defines a trivial conservation law of the second kind as one that obeys Eqs \eqref{defflux} and \eqref{defolver}.

\citet{Padhye96} claim that Eq.\ \eqref{defolver} --- which is equivalent to their eq.\ (31) --- is a ``generalized Bianchi identity''. To be a Bianchi identity, the term $\partial (\sqrt{{\cal G}}\rho_0 q) / \partial \tau$ would have to be expressible as a combination of $\Lambda_{(x^j)}$'s (and their derivatives) different from
\begin{align}
-\frac{\partial}{\partial a^l} \left( \frac{\varepsilon^{0lmn}}{\rho_0}\frac{\partial s}{\partial a^n} \frac{\partial x^j}{\partial a^m} \Lambda_{(x^j)} \right). \nonumber
\end{align}
They do not demonstrate that this is the case. Considering that Eq.\ \eqref{defolver} is a trivial conservation law, by definition it cannot be a Bianchi identity leading --- from Noether's second theorem --- to local gauge invariance, i.e.\ to a non-trivial conservation law.

The quantity
\begin{align}
\int d^3a \sqrt{{\cal G}} \rho_0 q = \int d^3x \sqrt{g} \rho q
\end{align}
is time-independent provided that the net flux $F^l$ across the integration limits vanishes, whether or not the equations of motion are satisfied.

\subsubsection{Relabeling is a symmetry --- but unrelated to Ertel's theorem}
As already mentioned, Noether's theorems have been invoked to associate particle-relabeling with Ertel's theorem \citep[][and others]{Padhye96,Salmon98}. Consider Noether's first theorem for the particle-like formulation in label representation. The action functional is written
\begin{align}
{\cal S}=\int d\tau \, d^3a \sqrt{{\cal G}} \rho_0 (K-I-\Phi),
\end{align}
with $K=g_{\mu\nu}u^\mu u^\nu /2 - 1/2$.

Following \citet{Padhye96}, a relabeling transformation will be interpreted as a passive transformation in label representation. In this context, an infinitesimal transformation of the labels $(a^1,a^2,a^3)$ is strictly equivalent to a ``coordinate'' transformation from $(a^1,a^2,a^3)$ to $(\hat{a}^1,\hat{a}^2,\hat{a}^3)$ at $\tau=0$, where
\begin{align}
\hat{a}^l = a^l + \vartheta^l,
\end{align}
with $\vartheta^l$ infinitesimally small. A passive variation of the action functional ${\cal S}$ under such relabeling is written
\begin{align}
\hat{\delta} {\cal S} = \int d\tau d^3a \left[ \hat{\delta}(\sqrt{{\cal G}} \rho_0)(K-I-\Phi) + \sqrt{{\cal G}} \rho_0 (\hat{\delta}K - \hat{\delta}I -\hat{\delta}\Phi) \right]. \label{labelds}
\end{align}
In this particle-like formulation, the dynamical fields are the spatial coordinates $x^i=x^i({\mathbf a},\tau)$. Note however that in general $x^i({\mathbf a},\tau=0) \ne a^i$ because the labels may be chosen independently of the actual coordinates, and that $g({\mathbf a},\tau=0)$ (the determinant of $g_{\mu\nu}$ at $x^0=\tau=0$ in the $x^\mu$ coordinates) is in general different from ${\cal G}$.

Under an infinitesimal relabeling, the coordinates $x^i({\mathbf a},\tau)$ becomes $\hat{x}^i(\hat{{\mathbf a}},\tau)$. The symbol $\hat{x}^i$ indicates a different functional form induced by the relabeling, however the value of the actual coordinate at a given point does not change under a relabeling: $x^i({\mathbf a},\tau)=\hat{x}^i(\hat{{\mathbf a}},\tau)=\hat{x}^i({\mathbf a},\tau)+(\partial x^i/\partial a^j) \, \vartheta^j$ to first order. The variation induced on the dynamical fields $x^i$ by a relabeling interpreted as a passive transformation is therefore given by
\begin{align}
\hat{\delta}x^i \equiv \hat{x}^i({\mathbf a},\tau)-x^i({\mathbf a},\tau)=-\frac{\partial x^i}{\partial a^j}\vartheta^j.
\end{align}

The mass of a given fluid particle, $dx^1dx^2dx^3 \sqrt{g} \rho  \equiv d^3x \sqrt{g} \rho$, is assumed to be preserved as time evolves. Therefore,
\begin{align}
d^3x \sqrt{g} \rho = d^3a \sqrt{{\cal G}} \rho_0.
\end{align}
At $\tau=0$, a volume element in coordinates $(x^1,x^2,x^3)$ is equal to a volume element in coordinates $(a^1,a^2,a^3)$: $d^3x \sqrt{g}(\tau=0) = d^3a \sqrt{{\cal G}}$. Therefore, $\rho_0({\mathbf a})=\rho({\mathbf a},\tau=0)$ at all spatial points (as mentioned previously).

In the particle-like formulation, mass and entropy are assumed conserved {\it a priori}. Therefore, the variation $\hat{\delta} s$ must vanish
\begin{align}
\hat{\delta} s=0, \label{cent}
\end{align}
and the variation $\hat{\delta} \rho$ must satisfy mass conservation. Recall that passive transformations are expressed in the original coordinates --- only the fields, external forcing, and metric terms change. Similarly, passive transformations in label representation induce changes to $\rho_0$ (as a scalar) and $\sqrt{{\cal G}}$ (as in Appendix \ref{pvs}), but leave $d^3a$ unchanged. The passive transformation of a mass element is $\hat{\delta}(d^3a \sqrt{{\cal G}}\rho_0)=\hat{\delta}(\sqrt{{\cal G}}\rho_0)d^3a=0$, implying that
\begin{align}
\hat{\delta}(\sqrt{{\cal G}}\rho_0)=0. \label{labeldm}
\end{align}
The mass preservation of all fluid particles also implies that the variation $\hat{\delta}(\rho\sqrt{g}d^3x)$ vanishes. Under a relabeling interpreted as a passive transformation, $\hat{\delta}(\sqrt{g})$ is expressed as $(\partial (\sqrt{g})/\partial x^i) \, \hat{\delta}x^i$ --- results from Appendix \ref{pvs} do not apply here because a relabeling is not a change of arbitrary coordinates interpreted as dynamical fields. Moreover, the variation $\hat{\delta}(d^3x)$ is expressed as $d^3\hat{x}({\mathbf a})-d^3x({\mathbf a})$, with $d^3\hat{x}({\mathbf a})=(1+\partial (\hat{\delta} x^i)/\partial x^i)d^3x$. This leads to $\hat{\delta}(d^3x)=(\partial (\hat{\delta} x^i)/\partial x^i)d^3x$. In order to preserve mass, the variation $\hat{\delta}\rho$ must therefore be expressed as
\begin{align}
\hat{\delta}\rho &=-\frac{\rho}{\sqrt{g}}\frac{\partial}{\partial x^i}(\sqrt{g} \hat{\delta}x^i), \nonumber \\
&= \frac{\rho}{\sqrt{g}}\frac{\partial}{\partial a^k} \left( \sqrt{g} \frac{\partial x^i}{\partial a^j} \vartheta^j \right) \frac{\partial a^k}{\partial x^i}.
\end{align}
Because $\rho$ is a scalar, it must also transform as $\hat{\delta} \rho = -(\partial \rho / \partial a^i) \vartheta^i$. This imposes a further constraint on relabeling:
\begin{align}
-\frac{\partial \rho}{\partial a^i}\vartheta^i = \frac{\rho}{\sqrt{g}}\frac{\partial}{\partial a^k} \left( \sqrt{g} \frac{\partial x^i}{\partial a^j} \vartheta^j \right) \frac{\partial a^k}{\partial x^i}. \label{cm3}
\end{align}

Under a relabeling transformation, since $K-I-\Phi$ is a scalar, the passive variation of $\sqrt{{\cal G}} \rho_0 (K-I-\Phi)$ is
\begin{align}
\hat{\delta} \left( \sqrt{{\cal G}} \rho_0 (K-I-\Phi) \right)&= \sqrt{{\cal G}} \rho_0 \hat{\delta}(K-I-\Phi) = -\sqrt{{\cal G}} \rho_0 \frac{\partial (K-I-\Phi)}{\partial a^l} \vartheta^l, \nonumber \\
&=-\frac{\partial}{\partial a^l} \left(  \sqrt{{\cal G}} \rho_0 \vartheta^l (K-I-\Phi) \right)+(K-I-\Phi)\frac{\partial}{\partial a^l} \left(  \sqrt{{\cal G}} \rho_0 \vartheta^l \right). \label{varl1}
\end{align}
The passive variation of $\sqrt{{\cal G}} \rho_0 (K-I-\Phi)$ may also be expressed as
\begin{align}
\hat{\delta} \left( \sqrt{{\cal G}} \rho_0 (K-I-\Phi) \right)&= \sqrt{{\cal G}}\rho_0 \left[ \frac{1}{2}u^\mu u^\nu \hat{\delta} g_{\mu\nu} + u_j \hat{\delta} u^j -\hat{\delta}I-\hat{\delta}\Phi  \right], \nonumber \\
&=\sqrt{{\cal G}}\rho_0 \vartheta^l \left[ -\frac{1}{2}u^\mu u^\nu \frac{\partial g_{\mu\nu}}{\partial a^l} -u_j \frac{\partial}{\partial \tau} \left( \frac{\partial x^j}{\partial a^l}\right)+ \frac{p}{\rho^2} \frac{\partial \rho}{\partial a^l} + \frac{\partial \Phi}{\partial a^l}\right] \nonumber \\
&\quad + \sqrt{{\cal G}}\rho_0 \vartheta^l T\frac{\partial s}{\partial a^l}, \nonumber \\
&=\sqrt{{\cal G}}\rho_0 \vartheta^l \frac{\partial x^j}{\partial a^l} \left[ \frac{\partial u_j}{\partial \tau}-\frac{1}{2}u^\mu u^\nu \frac{\partial g_{\mu\nu}}{\partial x^j} + \frac{1}{\rho}\frac{\partial p}{\partial x^j} + \frac{\partial \Phi}{\partial x^j}\right] \nonumber \\
& \quad - \sqrt{{\cal G}}\rho_0 \vartheta^l \frac{\partial}{\partial a^l}\left( \frac{p}{\rho} \right) + \sqrt{{\cal G}}\rho_0 \frac{\partial x^j}{\partial a^l} u_j \frac{\partial \vartheta^l}{\partial \tau} + \sqrt{{\cal G}}\rho_0 \vartheta^l T\frac{\partial s}{\partial a^l}\nonumber \\
& \quad -\frac{\partial}{\partial \tau} \left( \sqrt{{\cal G}}\rho_0 \vartheta^l \frac{\partial x^j}{\partial a^l} u_j \right), \nonumber \\
&=\sqrt{{\cal G}} \Lambda_{(x^j)} \hat{\delta} x^j + \frac{\partial}{\partial \tau} \left( \sqrt{{\cal G}} \rho_0 u_j \hat{\delta} x^j \right) + \sqrt{{\cal G}}\rho_0 \frac{\partial x^j}{\partial a^l} u_j \frac{\partial \vartheta^l}{\partial \tau} \nonumber \\
& \quad -\frac{\partial}{\partial a^l} \left( \sqrt{{\cal G}}\rho_0 \vartheta^l \frac{p}{\rho}  \right) + \frac{p}{\rho} \frac{\partial}{\partial a^l} \left( \sqrt{{\cal G}}\rho_0 \vartheta^l \right)+ \sqrt{{\cal G}}\rho_0 \vartheta^l T\frac{\partial s}{\partial a^l}, \label{varl2}
\end{align}
where $T$ is temperature. Equating Eqs \eqref{varl1} and \eqref{varl2}, one obtains Noether's first theorem under passive transformations in label representation:
\begin{align}
\frac{\partial}{\partial \tau} &\left( \sqrt{{\cal G}}\rho_0 u_j \hat{\delta} x^j \right) + \frac{\partial}{\partial a^l} \left( \sqrt{{\cal G}} \rho_0 \vartheta^l \left[ K-\Phi-I-\frac{p}{\rho} \right]  \right) \nonumber \\
&= - \sqrt{{\cal G}} \Lambda_{(x^j)} \hat{\delta} x^j + \left(K-\Phi-I-\frac{p}{\rho}  \right) \frac{\partial}{\partial a^l} \left( \sqrt{{\cal G}} \rho_0 \vartheta^l \right)-\sqrt{{\cal G}} \rho_0 \frac{\partial x^j}{\partial a^l} u_j \frac{\partial \vartheta^l}{\partial \tau} \nonumber \\
&\quad \; - T \sqrt{{\cal G}} \rho_0 \vartheta^l \frac{\partial s}{\partial a^l}. \label{nrl}
\end{align}
The three symmetry conditions
\begin{align}
\frac{\partial}{\partial a^l} \left( \sqrt{{\cal G}} \rho_0 \vartheta^l \right)&=0, \label{lc1} \\
\hat{\delta}s=-\frac{\partial s}{\partial a^l}\vartheta^l&=0, \\
\frac{\partial \vartheta^l}{\partial \tau}&=0 \label{lc3}
\end{align}
established by \citet{Padhye96} must be satisfied to obtain a conservation equation on-shell. The choice
\begin{align}
\vartheta^l = \frac{\varepsilon^{0lmn}}{\sqrt{{\cal G}}\rho_0} \frac{\partial s}{\partial a^m} \frac{\partial \xi}{\partial a^n}, \label{rls}
\end{align}
with $\xi({\mathbf a},\tau)=\chi(\tau)+\zeta({\mathbf a})$ and $\zeta$ any materially conserved infinitesimal function, satisfies the mass conservation condition \eqref{cm3} and leads to a symmetry transformation in label representation since it also satisfies the three symmetry conditions \eqref{lc1}--\eqref{lc3}.

Off-shell and once the symmetry conditions are explicitly satisfied, Eq.\ \eqref{nrl} reduces to
\begin{align}
\frac{\partial}{\partial a^0} \left( \sqrt{{\cal G}}\frac{\partial \zeta}{\partial a^l}A^{l0} \right) + \frac{\partial}{\partial a^m} \left( \sqrt{{\cal G}}\frac{\partial \zeta}{\partial a^l} A^{lm} \right) = \sqrt{{\cal G}}\frac{\partial \zeta}{\partial a^l}F^l, \label{rlnoether} \\
\sqrt{{\cal G}}\frac{\partial^2 \zeta}{\partial a^0 \partial a^l}A^{l0} + \sqrt{{\cal G}}\frac{\partial \zeta}{\partial a^l}F^l = \sqrt{{\cal G}}\frac{\partial \zeta}{\partial a^l}F^l
\end{align}
from the symmetry transformation given by Eq.\ \eqref{rls}. Here, Noether's first theorem is the statement that, provided suitable boundary conditions, the conserved charge associated with particle-relabeling is
\begin{align}
\int d^3a \sqrt{{\cal G}} \frac{\partial \zeta}{\partial a^l}A^{l0}=\int d^3a \, \zeta \frac{\partial (\sqrt{{\cal G}}A^{0l})}{\partial a^l} = \int d^3a \sqrt{{\cal G}} \rho_0 q \zeta = \int d^3x \sqrt{g} \rho q \zeta, \label{zetaq}
\end{align}
or that $\partial^2 \zeta/\partial a^0 \partial a^l=0$ off-shell, which is already known as equivalent to one of the symmetry conditions\footnote{A somewhat analogous situation exists in classical electromagnetism. Maxwell's equations exhibit a local gauge symmetry if the external electric 4-current is divergent-free, while Noether's second theorem leads to the conclusion that if a local gauge symmetry exists, then the external electric 4-current is divergent-free \citep[see for example][Appendix]{Brading02}.}. This is not Ertel's theorem.

Following the line of thought presented in \citet{Padhye96}, one may manipulate Eq.\ \eqref{rlnoether} and rewrite it as
\begin{align}
\zeta \left[ \frac{\partial}{\partial \tau}\left(\sqrt{{\cal G}}\rho_0 q\right) +\frac{\partial}{\partial a^l} \left(\sqrt{{\cal G}}F^l\right)\right]  = \frac{\partial}{\partial a^l} \left[ \sqrt{{\cal G}}\zeta F^l - \sqrt{{\cal G}}\frac{\partial \zeta}{\partial a^m}A^{ml} - \zeta \frac{\partial}{\partial \tau} \left(\sqrt{{\cal G}}A^{l0}\right) \right] \label{pm96}
\end{align}
after using $\partial \zeta / \partial \tau=0$. \citet{Padhye96}, working on-shell (i.e.\ $F^l=0$), integrate this equation over the labels, use the divergence theorem with appropriate boundary conditions, and conclude from the du Bois-Reymond lemma that potential vorticity is materially conserved {\it because} $\zeta$ is arbitrary. However, both the left-hand side and right-hand side of Eq.\ \eqref{pm96} vanish {\it identically}: the arbitrariness of $\zeta$, the du Bois-Reymond lemma, and therefore Noether's theorems have nothing to do with the fact that the material derivative of Ertel's potential vorticity vanishes on-shell. This result is a consequence of the trivial conservation law given by Eq.\ \eqref{defolver}.

The fact that Eq.\ \eqref{defolver} has not been recognized as a trivial conservation law has led to the unfounded association between the particle-relabeling symmetry and Ertel's potential vorticity conservation. Particle-relabeling is an actual symmetry of the dynamical system formulated in label representation, however the application of Noether's first theorem only leads to the consistency check of a symmetry condition, given that Ertel's theorem is a trivial conservation law obtained without assuming any symmetry. Noether's first theorem under a relabeling is a circular statement on the material conservation of $\zeta$, not a statement on the material conservation of $q$.

In sum, there is no contradiction between the covariant field theoretic formulation in arbitrary coordinates and the particle-like formulation in label representation: in both formulations, Ertel's theorem is a trivial conservation law of the second kind with no associated symmetry (particle-relabeling or other).

\section{Hamiltonian formulations}\label{ham}
The manifestly covariant Lagrangian field theory in arbitrary coordinates presented above uses Clebsch potentials as dynamical fields. In practice, it is often difficult to determine these potentials. The Hamiltonian formulation allows to choose more traditional dynamical fields, such as $u^i$. Moreover, if one is able to write the dynamical evolution of a physical system within a Hamiltonian formalism, one may take advantage of a vast body of general techniques to tackle specific problems, for instance non-linear stability theorems \citep[see e.g.][and references therein]{Shepherd90}. In addition and perhaps more importantly in the context of this paper, the theory of constrained Hamiltonians developed by \citet{Dirac50,Dirac64} and its classification of phase space constraints provide a systematic method to determine the existence of degrees of freedom associated with local gauge symmetries which are related to Noether's second theorem.

Hamiltonian formulations applied to fluid flows are reviewed for example in \citet{Shepherd90, Salmon98, Morrison98} and references therein.

In this section, canonical and non-canonical Hamiltonian formulations are developed with time-dependent constraints in arbitrary coordinates admissible in Newtonian fluid mechanics. As for the previous sections, results are valid under any geometric approximations, i.e.\ in curved (Riemannian) space as long as time intervals remain absolute. Dynamical approximations are not explicitly considered, but the methods described below could be applied under such approximations.

\subsection{Canonical Hamiltonian formulation with time-dependent constraints}
In a canonical Hamiltonian formulation, the ``velocities'' $\psi_{(p),0}$ of a dynamical field $\psi_{(p)}$ are replaced by ``momenta'' defined by
\begin{align}
\pi_{(\psi_{(p)})}&\equiv \frac{\partial (\sqrt{g}{\cal L})}{\partial (\psi_{(p),0})}, \nonumber \\
&=\sqrt{g}\frac{\partial {\cal L}}{\partial (\psi_{(p),0})}.
\end{align}
The phase space of the dynamical system consists of the $P$ dynamical fields $\psi_{(p)}$ and the $P$ conjugate momenta $\pi_{(\psi_{(p)})}$. In particular for the Lagrangian density \eqref{lag}, the momenta of the $4+2N$ scalar fields are
\begin{align}
\pi_{(\rho)}&=0, \\
\pi_{(\alpha)}&=-\sqrt{g}\rho, \\
\pi_{(\beta)}&=0, \\
\pi_{(s)}&=-\sqrt{g}\rho\beta, \\
\pi_{(\gamma_{(r)})}&=0, \\
\pi_{(\lambda_{(r)})}&=-\sqrt{g}\rho\gamma_{(r)}.
\end{align}
In principle, these relations are used to express the ``velocities'' in terms of $\psi_{(p)}$ and the corresponding momenta. However, when a Lagrangian density depends linearly on the ``velocities'', this is not possible and one finds contraints in phase space. This is the case of the Lagrangian density for geophysical fluid dynamics, where one therefore finds $P$ constraints $\phi_{(\psi_{(p)})}$ written as
\begin{align}
\phi_{(\rho)}&=\pi_{(\rho)} \approx 0, \\
\phi_{(\alpha)}&=\pi_{(\alpha)}+\sqrt{g}\rho  \approx 0, \\
\phi_{(\beta)}&=\pi_{(\beta)}  \approx 0, \\
\phi_{(s)}&=\pi_{(s)}+\sqrt{g}\rho\beta  \approx 0, \\
\phi_{(\gamma_{(r)})}&=\pi_{(\gamma_{(r)})}  \approx 0, \\
\phi_{(\lambda_{(r)})}&=\pi_{(\lambda_{(r)})}+\sqrt{g}\rho\gamma_{(r)}  \approx 0,
\end{align}
where the notation $\approx 0$ is explained later on. The Hamiltonian $H_C$ is
\begin{align}
H_C=\int d^3x \;{\cal H}_C,
\end{align}
where its density ${\cal H}_C$ is
\begin{align}
{\cal H}_C&=\sum_{p=1}^P \left( \pi_{(\psi_{(p)})} \psi_{(p),0} + {c'}_{(\psi_{(p)})} \phi_{(\psi_{(p)})}  \right) - \sqrt{g}{\cal L}, \nonumber \\
&=\sqrt{g}\rho\left( \frac{1}{2}h^{ij}u_i u_j+\Phi+I+g^{0i}u_i \right) + \sum_{p=1}^P c_{(\psi_{(p)})} \phi_{(\psi_{(p)})}. \label{hamiltonien}
\end{align}
The $c_{(\psi_{(p)})}$'s, where $c_{(\psi_{(p)})}={c'}_{(\psi_{(p)})}+\psi_{(p),0}$, are at this stage unknown functions to be determined. This Hamiltonian density incorporates the null constraints. Although it obeys the principle of Newtonian relativity, it is not manifestly invariant because time is singled out in the Hamiltonian formalism.

The equations of motion are obtained from the Poisson brackets\footnote{In this section, the fields and metric terms are considered at the same time $t$, therefore ${\mathbf x}$ and ${\mathbf y}$ are meant to represent two spatial --- not space-time --- points in the same coordinate system.}:
\begin{align}
\psi_{(p),0}({\mathbf x}) &= \left\{ \psi_{(p)}({\mathbf x}) , H_C \right\}, \label{hameq} \\
&= \int d^3y \left\{ \psi_{(p)}({\mathbf x}) , {\cal H}_C ({\mathbf y}) \right\},
\end{align}
and
\begin{align}
\pi_{(\psi_{(p)}),0}({\mathbf x}) = \int d^3y \left\{ \pi_{(\psi_{(p)})}({\mathbf x}) , {\cal H}_C ({\mathbf y}) \right\}.
\end{align}
The fundamental (canonical) Poisson brackets are
\begin{align}
\left\{ \psi_{(p)}({\mathbf x}) , \pi_{(\psi_{(q)})}({\mathbf y}) \right\} = -\left\{ \pi_{(\psi_{(p)})}({\mathbf x}) , \psi_{(q)}({\mathbf y}) \right\} = \delta_{(pq)}\delta^{(3)}({\mathbf x}-{\mathbf y}),
\end{align}
and zero otherwise.

Following the terminology and notation of \citet{Dirac64}, the $P$ conditions $\phi_{(\psi_{(p)})}\approx 0$ are called primary constraints. The symbol $\approx 0$ indicates that they are {\it weak} equations in the sense that Poisson brackets must be calculated before actually applying the constraints.

The primary constraints $\phi_{(\psi_{(p)})}\approx 0$ may not be the only constraints of the dynamical system. The conditions $\phi_{(\psi_{(p)})}\approx 0$ should remain valid at all times, which is equivalent to imposing that
\begin{align}
\phi_{(\psi_{(p)}),0}({\mathbf x}) \equiv \left\{ \phi_{(\psi_{(p)})}({\mathbf x}) , H_C \right\} + \left( \frac{\partial \phi_{(\psi_{(p)})}({\mathbf x})}{\partial t}  \right)_{\psi,\pi} \approx 0. \label{consco}
\end{align}
This relation means that the time derivative of a null constraint at a given point in space-time remains weakly zero. The last term before $\approx 0$ exists when a constraint has an explicit time dependence, for example involving the metric terms. The symbol $(\partial \cdot / \partial t)_{\psi,\pi}$ means that the time derivative is taken keeping the dynamical fields $\psi_{(p)}$ and their conjugate momenta $\pi_{(\psi_{(p)})}$ unchanged. If these $P$ equations are not equivalent to the already known primary constraints, they imply either new constraints (called secondary constraints) independent of the $c_{(\psi_{(p)})}$'s, or they determine the $c_{(\psi_{(p)})}$'s. When Eq.\ \eqref{consco} is applied successively on the constraints $\phi_{(\rho)}$, $\phi_{(\alpha)}$, $\phi_{(\beta)}$, $\phi_{(s)}$, $\phi_{(\gamma_{(r)})}$, and $\phi_{(\lambda_{(r)})}$, the following set of relations is obtained:
\begin{align}
c_{(\alpha)} +\beta c_{(s)} +\sum_{r=1}^N \gamma_{(r)}c_{(\lambda_{(r)})}&=-({\cal B}+g^{0i}u_i), \\
c_{(\rho)} &=-\frac{1}{\sqrt{g}} \left( \left( \sqrt{g} \rho u^i  \right)_{,i} + \rho (\sqrt{g})_{,0}\right), \\
c_{(s)} &= -u^i s_{,i}, \\
\beta c_{(\rho)} + \rho c_{(\beta)} &= -\frac{1}{\sqrt{g}}\left( \left( \sqrt{g} \rho u^i  \beta \right)_{,i} +\rho\beta (\sqrt{g})_{,0} \right) + \rho T , \\
c_{(\lambda_{(r)})} &= -u^i \lambda_{(r),i}, \\
\gamma_{(r)} c_{(\rho)} + \rho c_{(\gamma_{(r)})} &= - \frac{1}{\sqrt{g}} \left( \left( \sqrt{g} \rho u^i  \gamma_{(r)} \right)_{,i} +\rho \gamma_{(r)} (\sqrt{g})_{,0}\right),
\end{align}
where
\begin{align}
{\cal B}=\frac{1}{2}h^{ij}u_i u_j+\Phi+I+\frac{p}{\rho}
\end{align}
is the Bernoulli function. The above relations simply determine the $c_{(\psi_{(p)})}$'s:
\begin{align}
c_{(\rho)} &=-\frac{1}{\sqrt{g}} \left( \left( \sqrt{g} \rho u^i  \right)_{,i} + \rho (\sqrt{g})_{,0}\right), \\
c_{(\alpha)} &= -({\cal B}+g^{0i}u_i)+\beta u^i s_{,i}+\sum_{r=1}^N \gamma_{(r)} u^i \lambda_{(r),i}, \\
c_{(\beta)} &= - u^i \beta_{,i} +T , \\
c_{(s)} &= -u^i s_{,i}, \\
c_{(\gamma_{(r)})} &= -u^i \gamma_{(r),i}, \\
c_{(\lambda_{(r)})} &= -u^i \lambda_{(r),i}.
\end{align}
Consequently, there are no secondary constraints. The Hamiltonian $H_C$ is now fully determined and the equations of motion \eqref{hameq} are
\begin{align}
\rho_{,0}&=-\frac{1}{\sqrt{g}} \left( \left( \sqrt{g} \rho u^i  \right)_{,i} + \rho (\sqrt{g})_{,0}\right), \label{eqmotion1} \\
\alpha_{,0}&=-({\cal B}+g^{0i}u_i)+\beta u^i s_{,i}+\sum_{r=1}^N \gamma_{(r)} u^i \lambda_{(r),i}, \nonumber \\
&=-u^i \alpha_{,i}+\frac{1}{2}h^{ij}u_iu_j-I-\Phi-\frac{p}{\rho}, \\
\beta_{,0}&= -u^i \beta_{,i} +T, \\
s_{,0}&=-u^i s_{,i}, \label{eqmotions} \\
\gamma_{(r),0}&=-u^i \gamma_{(r),i}, \\
\lambda_{(r),0}&=-u^i \lambda_{(r),i} \label{eqmotion2}.
\end{align}
These are equivalent to eqs (71)--(76) in \citet{Zadra15}.

\subsubsection{Absence of local gauge symmetries} \label{ngs}
\citet{Dirac64} makes the distinction between first-class and second-class constraints. First-class constraints exist if and only if the determinant of the matrix $\mathsf{M}$ vanishes, where its matrix elements are
\begin{align}
\mathsf{M}_{(pq)}({\mathbf x},{\mathbf y})=\left\{ \phi_{(\psi_{(p)})}({\mathbf x}) , \phi_{(\psi_{(q)})}({\mathbf y})  \right\},
\end{align}
including all (primary and secondary) constraints. In this case, when dynamical fields are ordered as $(\rho,\alpha,\beta,s,\allowbreak\gamma_{(1)},\lambda_{(1)},...,\gamma_{(N)},\lambda_{(N)})$ and for $N=2$, this matrix reads
\begin{align}
\mathsf{M}({\mathbf x},{\mathbf y})= \sqrt{g} \; \delta^{(3)}({\mathbf x}-{\mathbf y})
\left( \begin{array}{cccccccc}
 0           & -1 & 0    & -\beta & 0    & -\gamma_{(1)} &  0   & -\gamma_{(2)} \\
 1           &  0 & 0    & 0      & 0    & 0             &  0   & 0             \\
 0           &  0 & 0    & -\rho  & 0    & 0             &  0   & 0             \\
\beta        &  0 & \rho & 0      & 0    & 0             &  0   & 0             \\
 0           &  0 & 0    & 0      & 0    & -\rho         &  0   & 0             \\
\gamma_{(1)} &  0 & 0    & 0      & \rho & 0             &  0   & 0             \\
 0           &  0 & 0    & 0      & 0    & 0             &  0   & -\rho         \\
\gamma_{(2)} &  0 & 0    & 0      & 0    & 0             & \rho & 0
\end{array} \right).
\end{align}
From the definition $\mathsf{M}({\mathbf x},{\mathbf y})\equiv \sqrt{g}\;\delta^{(3)}({\mathbf x}-{\mathbf y}) {\cal M}({\mathbf x})$, one may verify that
\begin{align}
\text{det}({\cal M})=\rho^{2(N+1)} \ne 0,
\end{align}
which holds for any positive integer $N$. There are consequently no first-class constraints in inviscid geophysical fluid dynamics when the governing equations are unapproximated or geometrically approximated. All constraints are of the second class. \citet{Dirac64} shows that local gauge symmetries exist only when there are first-class constraints. Therefore, there is no local gauge symmetry in the dynamical system presented in this paper.

When there is no local gauge symmetry, equations of motion are not under-determined. In this case, Noether's second theorem cannot be invoked because it only applies to under-determined systems.

The transformations provided by Eqs \eqref{lgt1}--\eqref{lgt2} are not local gauge transformations as $\epsilon$ is a constant. This is in contrast with classical electromagnetism, where the tensor $F_{\mu\nu}=A_{\mu:\nu}-A_{\nu:\mu}$ is invariant under the transformation $A_\mu \rightarrow A'_\mu=A_\mu+\psi_{,\mu}$, where the scalar $\psi$ is an arbitrary function of space-time. Such a freedom does not exist in geophysical fluid dynamics.

\subsection{Non-canonical Hamiltonian structure and symplectic form}
In Dirac's theory on generalized Hamiltonian systems, weak constraints become strong constraints, and Poisson brackets are replaced by Dirac brackets. 

In this case, once constraints become strong, the Hamiltonian density ${\cal H}_C$ given by Eq.\ \eqref{hamiltonien} becomes $\sqrt{g}(E^0-T^0_0)$. It is not in general the total energy density in all admissible coordinate systems. It reduces to the total energy density only in particular coordinate systems in which the components $T^0_i$ do not contribute to the total energy, as follows from Eq.\ \eqref{emc}.

A non-canonical Hamiltonian formulation may be obtained by constructing the Dirac brackets. One must first find the inverse of $\mathsf{M}$ --- which must be built with second-class constraints only --- from the following definition:
\begin{align}
\sum_{r=1}^P \int d^3 z \; {\mathsf{M}^{-1}}_{(pr)}({\mathbf x},{\mathbf z}) {\mathsf M}_{(rq)}({\mathbf z},{\mathbf y}) = \delta_{(pq)} \delta^{(3)}({\mathbf x}-{\mathbf y}).
\end{align}
For $N=2$, one obtains
\begin{align}
\mathsf{M}^{-1}({\mathbf x},{\mathbf y})=\frac{\delta^{(3)}({\mathbf x}-{\mathbf y})}{\sqrt{g}\rho}\left( \begin{array}{cccccccc}
 0    &  \rho         & 0     & 0 & 0            & 0 & 0            & 0 \\
-\rho &  0            & \beta & 0 & \gamma_{(1)} & 0 & \gamma_{(2)} & 0 \\
 0    & -\beta        & 0     & 1 & 0            & 0 & 0            & 0 \\
 0    &  0            & -1    & 0 & 0            & 0 & 0            & 0 \\
 0    & -\gamma_{(1)} & 0     & 0 & 0            & 1 & 0            & 0 \\
 0    &  0            & 0     & 0 & -1           & 0 & 0            & 0 \\
 0    & -\gamma_{(2)} & 0     & 0 & 0            & 0 & 0            & 1 \\
 0    &  0            & 0     & 0 & 0            & 0 & -1           & 0 \end{array} \right).
\end{align}
This is easily generalized to any positive integer $N$.

A Dirac bracket between $A$ and $B$ is defined by
\begin{align}
&\left[ A({\mathbf x}) , B({\mathbf y}) \right] \equiv \left\{ A({\mathbf x}) , B({\mathbf y}) \right\} \nonumber \\ &\quad - \sum_{p=1}^P \sum_{q=1}^P\int d^3z \int d^3w \left\{ A({\mathbf x}), \phi_{(\psi_{(p)})}({\mathbf w})\right\} {\mathsf{M}^{-1}}_{(pq)}({\mathbf w},{\mathbf z}) \left\{ \phi_{(\psi_{(q)})}({\mathbf z}), B({\mathbf y}) \right\} \label{defdb}
\end{align}
\citep{Dirac64}. This definition implies that the Dirac bracket between a second-class constraint and an arbitrary field $B({\mathbf y})$, written $\left[ \phi_{(\psi_{(p)})}({\mathbf x}) , B({\mathbf y}) \right]$, vanishes by construction. Therefore, when using Dirac brackets instead of Poisson brackets, second-class constraints become strong constraints; one may apply the second-class constraints from the start and there is no need to first calculate the brackets involving the second-class constraints.

The time evolution of a field $A({\mathbf x})$ at a point ${\mathbf x}$ is provided by
\begin{align}
A_{,0}({\mathbf x})= \left\{ A({\mathbf x}) , H_C \right\} + \left( \frac{\partial A({\mathbf x})}{\partial t} \right)_{\psi,\pi}. \label{tea}
\end{align}
The last term is non-vanishing only when an explicit time dependence of $A$ exists.

A way to take into account the explicit time dependence of a field is to follow \citet[][section 7.2]{Gitman90} and formally expand the phase space of dynamical variables to include the time itself and its conjugate momentum density $\pi_{(t)}$. In this extended phase space, the Poisson bracket of any function $A({\mathbf x})$ with $\pi_{(t)}({\mathbf y})$ is
\begin{align}
\left\{ A({\mathbf x}) , \pi_{(t)}({\mathbf y}) \right\} = \left( \frac{\partial A({\mathbf x})}{\partial t} \right)_{\psi,\pi}\delta^{(3)}({\mathbf x}-{\mathbf y}).
\end{align}
Consequently, the Poisson bracket of a function $A({\mathbf x})$ with $\pi_{(t)}({\mathbf y})$ is non-zero only if $A({\mathbf x})$ has an explicit time dependence. After defining a new Hamiltonian $H'$ as
\begin{align}
H'=H_C + \int d^3 x \; \pi_{(t)} ({\mathbf x}),
\end{align}
Eq.\ \eqref{tea} becomes
\begin{align}
A_{,0}({\mathbf x}) &= \left\{ A({\mathbf x}) , H' \right\}, \nonumber \\
&= \left[ A({\mathbf x}) , H' \right] \nonumber \\ &\quad + \sum_{p=1}^P \sum_{q=1}^P\int d^3z \int d^3w \left\{ A({\mathbf x}), \phi_{(\psi_{(p)})}({\mathbf w})\right\} {\mathsf{M}^{-1}}_{(pq)}({\mathbf w},{\mathbf z}) \left\{ \phi_{(\psi_{(q)})}({\mathbf z}), H' \right\}, \nonumber \\
&= \left[ A({\mathbf x}) , H' \right] + \sum_{p=1}^P \sum_{q=1}^P\int d^3z \int d^3w \left\{ A({\mathbf x}), \phi_{(\psi_{(p)})}({\mathbf w})\right\} {\mathsf{M}^{-1}}_{(pq)}({\mathbf w},{\mathbf z}) \phi_{(\psi_{(q)}),0}({\mathbf z}), \nonumber \\
&=\left[ A({\mathbf x}) , H' \right]
\end{align}
from Eqs \eqref{consco} and \eqref{defdb}. Moreover because the Dirac bracket between a second-class constraint and any field vanishes by construction, one may explicitly set $\phi_{(\psi_{(p)})}$ to zero in Eq.\ \eqref{hamiltonien} when using the non-canonical formalism. One therefore obtains the Hamiltonian $H$, where
\begin{align}
H=\int d^3x\left[ \sqrt{g}\rho\left( \frac{1}{2}h^{ij}u_i u_j+\Phi+I+g^{0i}u_i \right) +\pi_{(t)} \right]. \label{hamf}
\end{align}
The use of the non-canonical formalism (i.e.\ Dirac brackets instead of Poisson brackets) allows to replace $H'$ by $H$ in all equations. Therefore,
\begin{align}
A_{,0}({\mathbf x})= \left[ A({\mathbf x}) , H \right].
\end{align}

One must calculate the Dirac brackets among all the dynamical fields of the extended phase space. Note that the momenta $\pi_{(\psi_{(p)})}$ do not appear in the Hamiltonian \eqref{hamf}, as the constraints are now strong constraints. From the definition \eqref{defdb} and $\{ \psi_{(p)}({\mathbf x}),\psi_{(q)}({\mathbf y}) \}=0$, one may find among the $4+2N$ dynamical fields $\psi_{(p)}$
\begin{align}
\left[ \rho({\mathbf x}) , \alpha({\mathbf y}) \right]=-\left[ \alpha({\mathbf x}) , \rho({\mathbf y}) \right]&=(\sqrt{g})^{-1}\delta^{(3)}({\mathbf x}-{\mathbf y}), \label{drhoalpha} \\
\left[ \alpha({\mathbf x}) , \beta({\mathbf y}) \right]=-\left[ \beta({\mathbf x}) , \alpha({\mathbf y}) \right]&=(\sqrt{g}\rho)^{-1}\beta\delta^{(3)}({\mathbf x}-{\mathbf y}), \label{cld1} \\
\left[ \alpha({\mathbf x}) , \gamma_{(r)}({\mathbf y}) \right]=-\left[ \gamma_{(r)}({\mathbf x}) , \alpha({\mathbf y}) \right]&=(\sqrt{g}\rho)^{-1}\gamma_{(r)}\delta^{(3)}({\mathbf x}-{\mathbf y}), \\
\left[ \beta({\mathbf x}) , s({\mathbf y}) \right]=-\left[ s({\mathbf x}) , \beta({\mathbf y}) \right]&=(\sqrt{g}\rho)^{-1}\delta^{(3)}({\mathbf x}-{\mathbf y}), \\
\left[ \gamma_{(r)}({\mathbf x}) , \lambda_{(r)}({\mathbf y}) \right]=-\left[ \lambda_{(r)}({\mathbf x}) , \gamma_{(r)}({\mathbf y}) \right]&=(\sqrt{g}\rho)^{-1}\delta^{(3)}({\mathbf x}-{\mathbf y}), \label{cld4}
\end{align}
and zero otherwise. The only non-vanishing Dirac bracket between the dynamical fields $\psi_{(p)}$ and $\pi_{(t)}$ is
\begin{align}
\left[ \rho({\mathbf x}) , \pi_{(t)}({\mathbf y}) \right]=-\left[  \pi_{(t)}({\mathbf x}) , \rho({\mathbf y}) \right]=-\rho (\sqrt{g})^{-1} (\sqrt{g})_{,0}  \delta^{(3)}({\mathbf x}-{\mathbf y}). \label{drhop}
\end{align}
For a generic external field $\Psi$ (e.g.\ $g_{\mu\nu}$, $\sqrt{g}$, $\Phi$), one has
\begin{align}
\left[ \Psi({\mathbf x}) , \pi_{(t)}({\mathbf y}) \right]=-\left[  \pi_{(t)}({\mathbf x}) , \Psi({\mathbf y}) \right]=\Psi_{,0} \, \delta^{(3)}({\mathbf x}-{\mathbf y}).
\end{align}
The above equations imply
\begin{align}
\left[ \sqrt{g}({\mathbf x})\rho({\mathbf x}) , \pi_{(t)}({\mathbf y}) \right]=0.
\end{align}

The equations of motion with a possibly time-varying metric tensor are written as
\begin{align}
\psi_{(p),0} = \left[ \psi_{(p)} , H \right]. \label{hameqnc}
\end{align}
It may be verified that Eq.\ \eqref{hameqnc} reproduces Eqs \eqref{eqmotion1}--\eqref{eqmotion2}. The term $\pi_{(t)}$ in Eq.\ \eqref{hamf} may be omitted if one chooses a time-independent metric tensor.

The form of $H$ suggests that dynamical fields may be transformed from Clebsch potentials to tensor wind components. The Hamiltonian $H$ may be rewritten as
\begin{align}
H=\int d^3 x\left[\sqrt{g}\rho\left( \frac{1}{2}g_{ij}u^i u^j+\Phi_e+I \right)+\pi_{(t)}\right] \label{hamphys}
\end{align}
after some algebra and after defining an effective gravitational potential $\Phi_e\equiv\Phi-h_{00}/2$. From the relation
\begin{align}
u^i=h^{ij}\left(\alpha_{,j}+\beta s_{,j}+\sum_{r=1}^N \gamma_{(r)} \lambda_{(r),j}\right) + g^{0i},
\end{align}
one calculates the following Dirac brackets:
\begin{align}
\left[ \rho({\mathbf x}) , u^i({\mathbf y}) \right] &= (\sqrt{g}({\mathbf x}))^{-1} h^{ij}({\mathbf y})\frac{\partial}{\partial y^j}\left( \delta^{(3)}({\mathbf x}-{\mathbf y}) \right), \label{drhou} \\
\left[ u^i({\mathbf x}) , \rho({\mathbf y}) \right]&= -(\sqrt{g}({\mathbf y}))^{-1} h^{ij}({\mathbf x})\frac{\partial}{\partial x^j}\left( \delta^{(3)}({\mathbf x}-{\mathbf y}) \right), \\
\left[ s({\mathbf x}) , u^i({\mathbf y}) \right] &= -(\sqrt{g}\rho)^{-1} h^{ij} s_{,j} \delta^{(3)}({\mathbf x}-{\mathbf y}), \label{dsu} \\
\left[ u^i({\mathbf x}), s({\mathbf y}) \right]  &= (\sqrt{g}\rho)^{-1} h^{ij} s_{,j} \delta^{(3)}({\mathbf x}-{\mathbf y}), \\
\left[ u^i({\mathbf x}) , u^j({\mathbf y}) \right] &= \left( \sqrt{g}\rho  \right)^{-1}\left( h^{ik}{u^j}_{:k}-h^{jk}{u^i}_{:k} \right) \delta^{(3)}({\mathbf x}-{\mathbf y}) \label{buu}
\end{align}
(Eq.\ \eqref{buu} is demonstrated in Appendix \ref{abuu}). These are the only non-vanishing Dirac brackets among the dynamical fields $\rho$, $s$ and $u^i$ on which the Hamiltonian \eqref{hamphys} depends. In addition to these, one must also include Eq.\ \eqref{drhop} as well as
\begin{align}
\left[ u^i({\mathbf x}) , \pi_{(t)}({\mathbf y}) \right] &= \left( {h^{ij}}_{,0} \left( g_{j0}+g_{jk}u^k \right) + {g^{0i}}_{,0} \right) \delta^{(3)}({\mathbf x}-{\mathbf y}), \nonumber  \\
&=-h^{ij} g_{j\mu,0}u^\mu \delta^{(3)}({\mathbf x}-{\mathbf y}), \nonumber \\
&=-\left( \Gamma^i_{00} +\frac{1}{2} h^{ij} h_{00,j} +h^{ij} g_{jk,0}u^k \right) \delta^{(3)}({\mathbf x}-{\mathbf y}) \label{dup}
\end{align}
and
\begin{align}
\left[ t({\mathbf x}) , \pi_{(t)}({\mathbf y}) \right] &= \delta^{(3)}({\mathbf x}-{\mathbf y}) \label{dtp}
\end{align}
in the list of fundamental brackets. The expression $t({\mathbf x})$ is meant to represent the same time at each spatial point, not that time depends on space. The terms $\Gamma^\alpha_{\mu\nu}$ are Christoffel symbols of the second kind:
\begin{align}
\Gamma^\alpha_{\mu\nu}=\frac{1}{2} g^{\alpha\beta}\left( g_{\beta\mu,\nu}+g_{\beta\nu,\mu}-g_{\mu\nu,\beta}  \right).
\end{align}
This non-canonical Hamiltonian structure is a generalization to arbitrary coordinates of what is presented in, for example, \citet{Shepherd90} and \citet{Morrison98}.

The non-canonical Hamiltonian structure is therefore established in terms of observable and measurable fields (except $\pi_{(t)}$), and any admissible coordinate system may be used. The equations of motion are
\begin{align}
\rho_{,0} &= \left[ \rho , H \right], \label{eqfrho} \\
s_{,0} &= \left[ s , H \right], \label{eqfs} \\
{u^i}_{,0} &= \left[ u^i , H \right]. \label{eqfm}
\end{align}
Equations \eqref{eqfrho} and \eqref{eqfs} lead to \eqref{eqmotion1} and \eqref{eqmotions} respectively, while \eqref{eqfm} to
\begin{align}
{u^i}_{,0}=-u^j {u^i}_{,j} -\Gamma^i_{00} -2\Gamma^i_{j0}u^j - \Gamma^i_{jk}u^j u^k -h^{ij} \left( \Phi_{,j} + \frac{1}{\rho} p_{,j} \right) \label{momi}
\end{align}
(see Appendix \ref{amomi}). Equation \eqref{momi} is equivalent to eq.\ (40) in \citet{Charron14a} provided that viscosity is neglected.

Define a state vector $\eta$, for instance $\eta=(\rho,\alpha,\beta,s,\gamma_{(r)},\lambda_{(r)},t,\pi_{(t)})$ or $\eta=(\rho,s,u^1,u^2,u^3,t,\pi_{(t)})$ with $\eta_{(a)}$ representing each dynamical field. Define also an operator ${\cal J}({\mathbf x},{\mathbf y})$ with component $(a,b)$ as
\begin{align}
{\cal J}_{(ab)}({\mathbf x},{\mathbf y}) = \left[ \eta_{(a)}({\mathbf x}) , \eta_{(b)}({\mathbf y})  \right].
\end{align}
The non-zero components of this antisymmetric operator are provided by Eqs \eqref{drhoalpha}--\eqref{cld4}, \eqref{drhop}, \eqref{dtp}, or by Eqs \eqref{drhop}, \eqref{drhou}--\eqref{dtp} --- depending on the choice of dynamical variables.

The Dirac bracket of two fields $A({\mathbf x})$ and $B({\mathbf y})$ that depend on the dynamical fields is written
\begin{align}
\left[ A({\mathbf x}),B({\mathbf y}) \right] &= \sum_{a=1}^P \int d^3w\; \frac{\delta A({\mathbf x})}{\delta \eta_{(a)}(\mathbf w)} \left[ \eta_{(a)}(\mathbf w) , B({\mathbf y}) \right], \nonumber \\
&= \sum_{a=1}^P \sum_{b=1}^P \int d^3w \int d^3z\; \frac{\delta A({\mathbf x})}{\delta \eta_{(a)}(\mathbf w)} \left[ \eta_{(a)}(\mathbf w) , \eta_{(b)}(\mathbf z) \right] \frac{\delta B({\mathbf y})}{\delta \eta_{(b)}(\mathbf z)} , \nonumber \\
&= \sum_{a=1}^P \sum_{b=1}^P \int d^3w \int d^3z\; \frac{\delta A({\mathbf x})}{\delta \eta_{(a)}(\mathbf w)} {\cal J}_{(ab)}({\mathbf w},{\mathbf z}) \frac{\delta B({\mathbf y})}{\delta \eta_{(b)}(\mathbf z)},
\end{align}
where $P$ is now the total number of dynamical fields in the extended phase space ($6+2N$ or $7$, depending on the choice of dynamical fields). This is the symplectic form. The terms like $\delta F({\mathbf x})/\delta G({\mathbf y})$ are meant to represent functional derivatives, for example $\delta \eta_{(a)}({\mathbf x})/\delta \eta_{(b)}({\mathbf y})=\delta_{(ab)}\delta^{(3)}({\mathbf x}-{\mathbf y})$.

The time evolution of a field $A({\mathbf x})$ that depends on the dynamical fields is therefore written
\begin{align}
A_{,0}({\mathbf x})=[A({\mathbf x}),H] = \sum_{a=1}^P \sum_{b=1}^P \int d^3w\int d^3z\int d^3y\; \frac{\delta A({\mathbf x})}{\delta \eta_{(a)}(\mathbf w)} {\cal J}_{(ab)}({\mathbf w},{\mathbf z}) \frac{\delta {\cal H}({\mathbf y})}{\delta \eta_{(b)}(\mathbf z)}.
\end{align}
The equations of motion take the form
\begin{align}
\eta_{(a),0}({\mathbf x}) = \sum_{b=1}^P \int d^3z\int d^3y\; {\cal J}_{(ab)}({\mathbf x},{\mathbf z}) \frac{\delta {\cal H}({\mathbf y})}{\delta \eta_{(b)}(\mathbf z)}.
\end{align}

\subsection{Casimir invariants}
By definition, a Casimir invariant $C$ (sometimes called a distinguished functional) exists if its Dirac brackets with all the dynamical fields vanish:
\begin{align}
\left[ C , \eta_{(a)}(\mathbf x) \right] = 0 &= \left[ \eta_{(a)}(\mathbf x) , C \right], \nonumber \\
&=\sum_{b=1}^P \int d^3z\left[ \eta_{(a)}(\mathbf x) , \eta_{(b)}({\mathbf z}) \right] \frac{\delta C}{\delta \eta_{(b)}({\mathbf z})}, \nonumber \\
&=\sum_{b=1}^P \int d^3z \int d^3y \;{\cal J}_{(ab)}({\mathbf x},{\mathbf z}) \frac{\delta {\cal C}({\mathbf y})}{\delta \eta_{(b)}({\mathbf z})} \label{kernel}
\end{align}
for all $a$ and for all points ${\mathbf x}$ within the interior domain, where ${\cal C}$ is the density associated with the Casimir functional $C$, i.e.\ $C=\int d^3 y\; {\cal C}({\mathbf y})$. Equation \eqref{kernel} means that Casimir invariants exist if the kernel of the operator ${\cal J}({\mathbf x},{\mathbf y})$ is not an empty set. They are not absolute; their existence depends on the choice of dynamical variables. Obviously, because $C$ has vanishing Dirac brackets with all the dynamical fields, its Dirac bracket with the Hamiltonian $H$ also vanishes, and $C$ is a constant of the motion given suitable boundary conditions.

Suppose one chooses $(\rho,\alpha,\beta,s,\gamma_{(r)},\lambda_{(r)},t,\pi_{(t)})$ as the dynamical fields in the extended phase space. One may verify that the quantity
\begin{align}
F=\int_{\cal D} d^3x \sqrt{g}\rho {\cal F}(s,\gamma_{(1)},\lambda_{(1)},...,\gamma_{(N)},\lambda_{(N)}), \label{charge}
\end{align}
where ${\cal F}$ is any local function of $s$, $\gamma_{(r)}$ and $\lambda_{(r)}$, {\it is not} a Casimir invariant --- in particular, this includes total mass (${\cal F}=1$) and total entropy (${\cal F}=s$). Recall that the conserved charge $F$ is obtained from a non-trivial conservation law and is associated with an internal symmetry, as shown in sub-section \ref{mec}. However, one may verify that the quantity
\begin{align}
Q=\int_{\cal D} d^3x \sqrt{g}\rho q,
\end{align}
where $q$ is Ertel's potential vorticity, has vanishing Dirac brackets with all the dynamical fields, and {\it is} therefore a Casimir invariant. As shown in section \ref{secertel}, the conserved charge $Q$ is obtained from a trivial conservation law and is dissociated from any symmetry. This is the only known Casimir invariant (up to a trivial multiplicative constant) when using $(\rho,\alpha,\beta,s,\gamma_{(r)},\lambda_{(r)},t,\pi_{(t)})$ as dynamical fields. Notice that although
\begin{align}
R=\int_{\cal D} d^3x \sqrt{g}\rho\, {\cal Q}(q),
\end{align}
where ${\cal Q}(q)$ is any local function of $q$, is a constant of the motion, it {\it is not} in general a Casimir invariant. In this case,
\begin{align}
\left[ R,\alpha({\mathbf y})\right]={\cal Q}-q\frac{d{\cal Q}}{dq}+\frac{d^2{\cal Q}}{dq^2} \sum_{r=1}^N \frac{\varepsilon^{0ijk}}{\sqrt{g}\rho} s_{,i} q_{,j} \lambda_{(r),k} \gamma_{(r)}
\end{align}
for any point ${\mathbf y}$ located in the interior domain ${\cal D}$. This expression vanishes only for ${\cal Q}(q)=bq$ ($b$ a constant).

Suppose now that one chooses $(\rho,s,u^1,u^2,u^3,t,\pi_{(t)})$ instead as the dynamical fields. It may be verified that the quantity
\begin{align}
C=\int_{\cal D} d^3x \sqrt{g}\rho \, {\cal C}_1(q,s),
\end{align}
where ${\cal C}_1(q,s)$ is any local function of Ertel's potential vorticity $q$ and specific entropy $s$, has vanishing Dirac brackets with the seven dynamical fields and {\it is} a Casimir invariant. In this case, no explicit or visible symmetries are associated with the time invariance of $C$. In particular, total mass (${\cal C}_1(q,s)=1$) and total entropy (${\cal C}_1(q,s)=s$) become Casimir invariants.

A trivial conservation law of the second kind --- that is, Ertel's theorem --- translates into a Casimir invariant in the non-canonical Hamiltonian formulation. This is true for the two choices of dynamical variables investigated here. It is however unknown to the authors of this paper whether this result may be generalized to any trivial conservation law of the second kind, or if it is a mere coincidence. It is interesting to note that trivial conservation laws and Casimir invariants are both dissociated from symmetries.

\section{Summary and conclusions}\label{conclu}
Starting from a manifestly invariant Lagrangian density for geophysical fluids --- expressed as a function of fluid density, specific entropy, other Clebsch potentials, the metric tensor and an external gravitational potential --- some symmetry properties of the equations of motion were investigated in the context of arbitrary coordinates. With that objective in mind, a review of Noether's first theorem for field transformations as well as coordinate transformations was presented. Using Noether's first theorem, it was shown that mass and entropy conservation laws are associated with internal symmetries under an active transformation of Clebsch fields. In addition, it was confirmed that momentum and energy conservation laws are associated with space-time symmetries of the external forcing and metric tensor. These results were presented as examples of applications of Noether's first theorem leading to non-trivial conservation laws. A non-trivial conservation law requires the existence of a symmetry, and is valid on-shell only.

In contrast, a trivial conservation law of the second kind is an off-shell mathematical identity obtained independently of the equations of motion and {\it a fortiori} of any assumed symmetry of these equations. Its 4-current is expressed as the covariant divergence of an antisymmetric second-rank tensor. In this paper, it was demonstrated that Ertel's theorem satisfies this condition and is therefore a trivial conservation law of the second kind. This leads to the dissociation of Ertel's theorem from symmetries. Consequently, the association of Ertel's theorem with the particle-relabeling symmetry made by several authors is unfounded.

Relabeling is a symmetry transformation in label representation parameterized by an arbitrary passive tracer $\zeta$. The associated conserved charge is $\int d^3x \sqrt{g} \rho q \zeta$. Given that $\int d^3x \sqrt{g}\rho q$ was shown to be trivially conserved (independently of relabeling or any symmetry transformation), it follows that the conservation law associated with relabeling is simply equivalent to the material conservation of $\zeta$ (which does nothing but confirm the prior assumption that $\zeta$ was a passive tracer).

Non-canonical Hamiltonian structures were derived in arbitrary coordinates from a canonical Hamiltonian formalism with weak constraints. In particular, a time-dependent metric tensor is accounted for by formally extending the phase space of dynamical fields to include time itself and its conjugate momentum. Using Dirac's theory for constrained Hamiltonian systems, it was shown that inviscid, but otherwise unapproximated geophysical fluid dynamics is not invariant under any local gauge transformation, therefore its equations of motion are not under-determined and Noether's second theorem does not apply. This remains true for geometrically approximated equations --- for instance, when the thin-shell approximation is used.

Here, it was shown that the conservation of Ertel's potential vorticity not only is a trivial law, but also translates into a Casimir invariant. Whether this is a coincidence or a general property of trivial conservation laws of the second kind remains unknown to the authors of this paper.
%


\begin{appendices}
\section{Useful identities}
\subsection{Mass-momentum tensor} \label{attt}
The mass-momentum tensor is related to the derivative of the Lagrangian density with respect to the metric tensor:
\begin{align}
t_{\mu\nu} &\equiv -\frac{2}{\sqrt{g}}\frac{\partial (\sqrt{g}{\cal L})}{\partial g^{\mu\nu}} \nonumber \\
&= -2 \left( \frac{{\cal L}}{\sqrt{g}} \frac{\partial (\sqrt{g})}{\partial g^{\mu\nu}} + \frac{\partial {\cal L}}{\partial g^{\mu\nu}}  \right).
\end{align}
The derivative of $\sqrt{g}$ with respect to $x^\alpha$ is written
\begin{align}
(\sqrt{g})_{,\alpha} &\equiv \frac{\partial (\sqrt{g})}{\partial g^{\mu\nu}} {g^{\mu\nu}}_{,\alpha}, \nonumber \\
&=\sqrt{g} \; \Gamma^\beta_{\alpha\beta}=\frac{1}{2}\sqrt{g}\; g^{\mu\nu}g_{\mu\nu,\alpha}, \nonumber \\
&=-\frac{1}{2}\sqrt{g}\; g_{\mu\nu}{g^{\mu\nu}}_{,\alpha}, \nonumber \\
&=-\frac{1}{2}\sqrt{g}\; (h_{\mu\nu}+ \delta^0_\mu\delta^0_\nu){g^{\mu\nu}}_{,\alpha}, \nonumber \\
&=-\frac{1}{2}\sqrt{g}\; h_{\mu\nu} {g^{\mu\nu}}_{,\alpha}
\end{align}
from eq.\ (22) in \citet{Charron14a} and $g^{00}=1$. The term $\Gamma^\mu_{\alpha\beta}=g^{\mu\nu}(g_{\nu\alpha,\beta}+g_{\nu\beta,\alpha}-g_{\alpha\beta,\nu})/2$ is a Christoffel symbol of the second kind. This implies that
\begin{align}
\frac{\partial (\sqrt{g})}{\partial g^{\mu\nu}} = -\frac{1}{2}\sqrt{g}\; h_{\mu\nu}
\end{align}
and that
\begin{align}
-2\frac{{\cal L}}{\sqrt{g}} \frac{\partial (\sqrt{g})}{\partial g^{\mu\nu}}= h_{\mu\nu}p +h_{\mu\nu}\rho\Lambda_{(\rho)},
\end{align}
from eq.\ (65) in \citet{Zadra15}. Note however that this relation does not hold for $\mu=\nu=0$ because $g^{00}$ is a constant.

The term $\partial {\cal L} / \partial g^{\mu\nu}$ is most directly obtained from the Lagrangian density \eqref{lag} when symmetrically rearranged, while making use of
\begin{align}
u^0&=1=g^{00} \\
v_i&=u_i \\
h^{\mu\nu}v_\mu v_\nu &= (g^{ij}-g^{0i}g^{j0})u_i u_j.
\end{align}
The Lagrangian density is written
\begin{align}
{\cal L}=-\frac{\rho}{2} \left( g^{ij}u_iu_j-g^{0i}g^{j0}u_iu_j +g^{0i}u_i+g^{i0}u_i +2I+2\Phi+2v_0 \right).
\end{align}
It is seen that
\begin{align}
-2\frac{\partial {\cal L}}{\partial g^{ij}}=\rho u_i u_j,
\end{align}
and that
\begin{align}
-2\frac{\partial {\cal L}}{\partial g^{0i}}&=\rho \left( 1-g^{j0}u_j  \right)u_i, \nonumber \\
&=\rho \left( 1-g^{\mu 0}u_\mu +g^{00}u_0 \right)u_i, \nonumber \\
&=\rho \left( 1-u^0 +u_0 \right)u_i, \nonumber \\
&=\rho u_0 u_i.
\end{align}
After defining the covariant mass-momentum tensor as $T_{\mu\nu}=\rho u_\mu u_\nu +h_{\mu\nu}p$, the tensor $t_{\mu\nu}$ is therefore written
\begin{align}
t_{\mu\nu}=T_{\mu\nu}+h_{\mu\nu}\rho\Lambda_{(\rho)} \label{mmt}
\end{align}
except when $\mu=\nu=0$. However, as a consequence of $\tilde{\delta} g^{00}$ being zero, the tensor component $t_{00}$ never appears in Eq.\ \eqref{delta1}. One may therefore define it arbitrarily. Equation \eqref{mmt} may then be used without restrictions even when $\mu=\nu=0$.

\subsection{Passive variation of $\sqrt{g}$}\label{pvs}
From the transformation
\begin{align}
g_{\mu\nu}(x)=\frac{\partial \tilde{x}^\alpha}{\partial x^\mu} \frac{\partial \tilde{x}^\beta}{\partial x^\nu}\tilde{g}_{\alpha\beta}(\tilde{x}), \label{gt}
\end{align}
one deduces
\begin{align}
\sqrt{\tilde{g}}(\tilde{x})=J^{-1}\sqrt{g}(x),
\end{align}
where $g(x)$ and $\tilde{g}(\tilde{x})$ are the determinant of the covariant metric tensor in the original and transformed coordinate systems, respectively. The quantity $J$ is the Jacobian of the transformation. For an infinitesimal transformation $\tilde{x}^\mu=x^\mu+\epsilon^\mu$,
\begin{align}
J&=\left| \begin{array}{cccc}
1+{\epsilon^0}_{,0} &   {\epsilon^1}_{,0} &   {\epsilon^2}_{,0} &   {\epsilon^3}_{,0} \\
  {\epsilon^0}_{,1} & 1+{\epsilon^1}_{,1} &   {\epsilon^2}_{,1} &   {\epsilon^3}_{,1} \\
  {\epsilon^0}_{,2} &   {\epsilon^1}_{,2} & 1+{\epsilon^2}_{,2} &   {\epsilon^3}_{,2} \\
  {\epsilon^0}_{,3} &   {\epsilon^1}_{,3} &   {\epsilon^2}_{,3} & 1+{\epsilon^3}_{,3} \end{array}\right|, \nonumber \\
&=1+{\epsilon^\mu}_{,\mu}
\end{align}
to first order. Therefore, $J^{-1}=1-{\epsilon^\mu}_{,\mu}$ to first order. After writing
\begin{align}
\sqrt{\tilde{g}}(\tilde{x})&=\sqrt{\tilde{g}}(x)+(\sqrt{\tilde{g}}(x))_{,\mu}\epsilon^\mu, \nonumber \\
&=\sqrt{\tilde{g}}(x)+(\sqrt{g}(x))_{,\mu}\epsilon^\mu, \nonumber \\
&=\sqrt{g}(x)-\sqrt{g}(x) {\epsilon^\mu}_{,\mu}
\end{align}
to first order, and defining
\begin{align}
\tilde{\delta} (\sqrt{g}) \equiv \sqrt{\tilde{g}}(x)-\sqrt{g}(x),
\end{align}
one obtains
\begin{align}
\tilde{\delta} (\sqrt{g}) = - (\sqrt{g} \epsilon^\mu)_{,\mu} = -\sqrt{g}{\epsilon^\mu}_{:\mu}.
\end{align}

\subsection{Passive variation of $g^{\mu\nu}$}\label{kf}
From the transformation
\begin{align}
\tilde{g}^{\mu\nu}(\tilde{x})=\frac{\partial \tilde{x}^\mu}{\partial x^\alpha} \frac{\partial \tilde{x}^\nu}{\partial x^\beta}g^{\alpha\beta}(x)
\end{align}
and the infinitesimal transformation $\tilde{x}^\mu=x^\mu+\epsilon^\mu$, one may write to first order
\begin{align}
\tilde{g}^{\mu\nu}(\tilde{x})&=\tilde{g}^{\mu\nu}(x) + (\tilde{g}^{\mu\nu}(x))_{,\beta} \epsilon^\beta, \nonumber \\
&=\tilde{g}^{\mu\nu}(x) + (g^{\mu\nu}(x))_{,\beta} \epsilon^\beta, \nonumber \\
&=g^{\mu\nu}(x) + g^{\mu\alpha}(x) {\epsilon^\nu}_{,\alpha} + g^{\alpha\nu}(x) {\epsilon^\mu}_{,\alpha}.
\end{align}
After defining $\tilde{\delta} g^{\mu\nu} \equiv \tilde{g}^{\mu\nu}(x)-g^{\mu\nu}(x)$, one obtains
\begin{align}
\tilde{\delta} g^{\mu\nu}&=g^{\alpha\nu} {\epsilon^\mu}_{,\alpha}+g^{\mu\alpha} {\epsilon^\nu}_{,\alpha}-{g^{\mu\nu}}_{,\beta} \epsilon^\beta, \nonumber \\
&=g^{\alpha\nu} {\epsilon^\mu}_{:\alpha}+g^{\mu\alpha} {\epsilon^\nu}_{:\alpha} -g^{\alpha\nu} \Gamma^\mu_{\alpha\beta}\epsilon^\beta - g^{\mu\alpha} \Gamma^\nu_{\alpha\beta}\epsilon^\beta + g^{\alpha\nu} \Gamma^\mu_{\alpha\beta}\epsilon^\beta + g^{\mu\alpha} \Gamma^\nu_{\alpha\beta}\epsilon^\beta, \nonumber \\
&=\epsilon^{\mu:\nu}+\epsilon^{\nu:\mu} \label{killing}
\end{align}
from the definition of covariant derivatives for first and second rank tensors, and the fact that ${g^{\mu\nu}}_{:\beta}$ vanishes. Relation \eqref{killing} is compatible with $\tilde{\delta} g^{00}=0$ because $\epsilon^0$ is a constant.

If a symmetry exists such that $\tilde{\delta} g^{\mu\nu}=0$, then $\epsilon^{\mu:\nu}=-\epsilon^{\nu:\mu}$, or equivalently $\epsilon_{\mu:\nu}=-\epsilon_{\nu:\mu}$ after contractions with the covariant metric tensor.

\subsection{Three tensor identities}
\begin{enumerate}
\item Consider an antisymmetric tensor $F^{\mu\nu}=-F^{\nu\mu}$. It will be shown that the scalar ${F^{\mu\nu}}_{:\nu:\mu}$ vanishes in a Riemannian space. The term $\sqrt{g}{F^{\mu\nu}}_{:\nu}$ is written
\begin{align}
\sqrt{g}{F^{\mu\nu}}_{:\nu}&=\sqrt{g}{F^{\mu\nu}}_{,\nu}+\sqrt{g}\Gamma^\mu_{\alpha\nu}F^{\alpha\nu}+\sqrt{g}\Gamma^\nu_{\alpha\nu}F^{\mu\alpha}, \nonumber \\
&=\sqrt{g}{F^{\mu\nu}}_{,\nu}+(\sqrt{g})_{,\nu}F^{\mu\nu}, \nonumber \\
&=(\sqrt{g}F^{\mu\nu})_{,\nu}. \label{ii3}
\end{align}
Moreover,
\begin{align}
{F^{\mu\nu}}_{:\nu:\mu} &= (\sqrt{g})^{-1}(\sqrt{g}{F^{\mu\nu}}_{:\nu})_{,\mu}, \nonumber \\
&=(\sqrt{g})^{-1}(\sqrt{g}F^{\mu\nu})_{,\nu,\mu}, \nonumber \\
&=0 \label{i3}
\end{align}
from Eq.\ \eqref{ii3}, the antisymmetry of $F^{\mu\nu}$, and the commutativity of ordinary derivatives.
\item The intrinsic derivative of the gradient of a scalar $f$ is written
\begin{align}
\frac{D}{Dt}(f_{,\mu})\equiv u^\nu f_{,\mu:\nu}=\dot{f}_{,\mu}-f_{,\nu} {u^\nu}_{:\mu}, \label{i1}
\end{align}
where $\dot{f}\equiv u^\nu f_{,\nu}$.
\item One may verify the identity
\begin{align}
\varepsilon^{\alpha\mu\nu\sigma}{A^\beta}_{\alpha} +  \varepsilon^{\beta\alpha\nu\sigma}{A^\mu}_{\alpha} + \varepsilon^{\beta\mu\alpha\sigma}{A^\nu}_{\alpha} + \varepsilon^{\beta\mu\nu\alpha}{A^\sigma}_{\alpha} =\varepsilon^{\beta\mu\nu\sigma}{A^\alpha}_{\alpha},
\end{align}
contract it with $\delta^0_\beta$, and replace ${A^\mu}_{\alpha}$ by ${u^\mu}_{:\alpha}$ to obtain
\begin{align}
\varepsilon^{0\alpha\nu\sigma}{u^\mu}_{:\alpha} + \varepsilon^{0\mu\alpha\sigma}{u^\nu}_{:\alpha} + \varepsilon^{0\mu\nu\alpha}{u^\sigma}_{:\alpha}=\varepsilon^{0\mu\nu\sigma}{u^\alpha}_{:\alpha}. \label{i2}
\end{align}
This holds because ${u^0}_{:\alpha}=0$ (recall that $u^0=1$). This is a tensor identity when multiplied by $(\sqrt{g})^{-1}$.
\end{enumerate}

\section{Second demonstration of Ertel's theorem as a trivial conservation law} \label{trivobs}
In this Appendix, Ertel's theorem expressed with observable dynamical fields is shown to be a trivial conservation law of the second kind. From the definition of Ertel's potential vorticity,
\begin{align}
\rho u^\alpha q&=\frac{\varepsilon^{0\mu\nu\sigma}}{\sqrt{g}}u_{\nu:\mu}s_{,\sigma} u^\alpha, \nonumber \\
&= \left( \frac{\varepsilon^{0\mu\nu\sigma}}{\sqrt{g}} u_\nu s_{,\sigma} u^\alpha  \right)_{:\mu} - \frac{\varepsilon^{0\mu\nu\sigma}}{\sqrt{g}} u_\nu s_{,\sigma} {u^\alpha}_{:\mu}, \nonumber \\
&= \left( u^\alpha B^\mu \right)_{:\mu} + \frac{1}{\sqrt{g}} u_\nu s_{,\sigma} \left( \varepsilon^{0\alpha\mu\sigma} {u^\nu}_{:\mu} + \varepsilon^{0\alpha\nu\mu} {u^\sigma}_{:\mu} - \varepsilon^{0\alpha\nu\sigma} {u^\mu}_{:\mu}\right), \nonumber \\
&= \left( u^\alpha B^\mu - u^\mu B^\alpha + \frac{\varepsilon^{0\alpha\mu\sigma}}{\sqrt{g}} \left[ s_{,\sigma} K + u_\sigma \frac{\Lambda_{(\beta)}}{\rho}  \right] \right)_{:\mu} -\omega^\alpha \frac{\Lambda_{(\beta)}}{\rho} \nonumber \\
&\qquad + u^\mu {B^\alpha}_{:\mu}-\frac{\varepsilon^{0\alpha\nu\mu}}{\sqrt{g}} u_\nu u^\sigma s_{,\sigma:\mu} \nonumber
\end{align}
after using Eqs \eqref{i2}, \eqref{ls}, \eqref{omeg}, \eqref{bbb} and $K=u_\nu u^\nu /2-1/2$. From
\begin{align}
u^\mu {B^\alpha}_{:\mu}-\frac{\varepsilon^{0\alpha\nu\mu}}{\sqrt{g}} u_\nu u^\sigma s_{,\sigma:\mu} &= \frac{\varepsilon^{0\alpha\mu\sigma}}{\sqrt{g}} u^\beta u_{\mu:\beta} s_{,\sigma}, \nonumber \\
&= \frac{\varepsilon^{0\alpha\mu\sigma}}{\sqrt{g}} \frac{(\Lambda_\mu-u_\mu \Lambda^0)}{\rho}s_{,\sigma} -\left( \frac{\varepsilon^{0\alpha\mu\sigma}}{\sqrt{g}} s_{,\sigma} \Phi \right)_{:\mu} \nonumber \\
&\qquad -\frac{\varepsilon^{0\alpha\mu\sigma}}{\sqrt{g}} \frac{p_{,\mu}}{\rho} s_{,\sigma}, \nonumber \\
&= \frac{\varepsilon^{0\alpha\mu\sigma}}{\sqrt{g}} \frac{(\Lambda_\mu-u_\mu \Lambda^0)}{\rho}s_{,\sigma} -\left( \frac{\varepsilon^{0\alpha\mu\sigma}}{\sqrt{g}} s_{,\sigma} \Phi \right)_{:\mu} \nonumber \\
&\qquad -\left( \frac{\varepsilon^{0\alpha\mu\sigma}}{\sqrt{g}} (I+p/\rho) s_{,\sigma} \right)_{:\mu},
\end{align}
one finally gets
\begin{align}
\rho u^\alpha q&=\left( u^\alpha B^\mu - u^\mu B^\alpha + \frac{\varepsilon^{0\alpha\mu\sigma}}{\sqrt{g}} \left[ s_{,\sigma} (K-\Phi-I-p/\rho) + u_\sigma \frac{\Lambda_{(\beta)}}{\rho}  \right] \right)_{:\mu} \nonumber \\
&\qquad -\omega^\alpha \frac{\Lambda_{(\beta)}}{\rho} + \frac{\varepsilon^{0\alpha\mu\sigma}}{\sqrt{g}} \frac{(\Lambda_\mu-u_\mu \Lambda^0)s_{,\sigma}}{\rho},
\end{align}
which proves Eq.\ \eqref{curqobs}.

\section{Explicit expression for $\tilde{\delta}{\cal S}_{KM}$} \label{secondt}
The term $\tilde{\delta}{\cal S}_{KM}$ is defined as
\begin{align}
\tilde{\delta}{\cal S}_{KM}=\int_{\cal D} d^4x \sqrt{g} \rho u^\mu v_\nu {\epsilon^\nu}_{:\mu}.
\end{align}
From Eq.\ \eqref{dcontra}, $v_\nu {\epsilon^\nu}_{:\mu}$ may be expressed as $-\tilde{\delta}v_\mu-\epsilon^\nu v_{\mu:\nu}$, and one may rewrite $\tilde{\delta}{\cal S}_{KM}$ as
\begin{align}
\tilde{\delta}{\cal S}_{KM}&=-\int_{\cal D} d^4x \sqrt{g} \rho u^\mu \tilde{\delta}v_\mu-\int_{\cal D} d^4x \sqrt{g} \rho \epsilon^\nu u^\mu v_{\mu:\nu}, \\
&=-\tilde{\delta}{\cal S}_{KD} - \int_{\cal D} d^4x \sqrt{g} \rho \epsilon^\nu u^\mu v_{\mu:\nu}.
\end{align}
In addition,
\begin{align}
\sqrt{g}\rho \epsilon^\nu u^\mu v_{\mu:\nu}&=\sqrt{g}\rho \epsilon^\nu (h^{\mu\alpha}v_\alpha+g^{0\mu} )v_{\mu:\nu}, \\
&=\sqrt{g}\rho \epsilon^\nu \left( \frac{1}{2} h^{\mu\alpha} v_\mu v_\alpha +g^{0\mu} v_\mu  \right)_{:\nu}, \\
&=\sqrt{g}\left( \rho \epsilon^\nu \left[ \frac{1}{2} h^{\mu\alpha} v_\mu v_\alpha +g^{0\mu} v_\mu  \right] \right)_{:\nu}, \\
&=\left( \sqrt{g}\rho \epsilon^\nu \left[ \frac{1}{2} h^{\mu\alpha} v_\mu v_\alpha +g^{0\mu} v_\mu  \right] \right)_{,\nu}
\end{align}
since $(\rho \epsilon^\nu)_{:\nu}=0$ from Eq.\ \eqref{passiveq}. Therefore,
\begin{align}
\int_{\cal D} d^4x \sqrt{g} \rho \epsilon^\nu u^\mu v_{\mu:\nu}&=\oint_{\partial {\cal D}} dS_\nu \rho \epsilon^\nu \left( \frac{1}{2} h^{\mu\alpha} v_\mu v_\alpha +g^{0\mu} v_\mu  \right), \\
&=0
\end{align}
given that $\epsilon^\nu$ vanishes on the contour $\partial {\cal D}$. This leads to $\tilde{\delta}{\cal S}_{KM}=-\tilde{\delta}{\cal S}_{KD}$, as expected.

\section{Demonstration of Eq.\ \eqref{buu}} \label{abuu}
First, the Dirac brackets of the Clebsch potentials with $u_j$ are calculated from Eqs \eqref{cld1}--\eqref{cld4} and from $u_j=\alpha_{,j}+\beta s_{,j}+\sum_{r=1}^N \gamma_{(r)} \lambda_{(r),j}$:
\begin{align}
\left[ \alpha({\mathbf x}) , u_j({\mathbf y}) \right] &= (\rho({\mathbf y})\sqrt{g}({\mathbf y}))^{-1} \left( \beta({\mathbf y}) \frac{\partial s({\mathbf y})}{\partial y^j} + \sum_{r=1}^N \gamma_{(r)}({\mathbf y}) \frac{\partial \lambda_{(r)}({\mathbf y})}{\partial y^j} \right) \delta^{(3)}({\mathbf x}-{\mathbf y}),\nonumber  \\
&= (\rho({\mathbf y})\sqrt{g}({\mathbf y}))^{-1} \left( \beta({\mathbf x}) \frac{\partial s({\mathbf y})}{\partial y^j} + \sum_{r=1}^N \gamma_{(r)}({\mathbf x}) \frac{\partial \lambda_{(r)}({\mathbf y})}{\partial y^j} \right) \delta^{(3)}({\mathbf x}-{\mathbf y}), \\
\left[ \beta({\mathbf x}) , u_j({\mathbf y}) \right] &= -\frac{\partial}{\partial y^j} \left( (\rho({\mathbf y})\sqrt{g}({\mathbf y}))^{-1} \beta({\mathbf y}) \delta^{(3)}({\mathbf x}-{\mathbf y}) \right) \nonumber \\
&\quad + \beta({\mathbf y}) \frac{\partial}{\partial y^j} \left((\rho({\mathbf y})\sqrt{g}({\mathbf y}))^{-1} \delta^{(3)}({\mathbf x}-{\mathbf y}) \right), \nonumber \\
&= -(\rho({\mathbf y})\sqrt{g}({\mathbf y}))^{-1} \frac{\partial \beta({\mathbf y})}{\partial y^j} \delta^{(3)}({\mathbf x}-{\mathbf y}), \\
\left[ s({\mathbf x}) , u_j({\mathbf y}) \right] &= -(\rho({\mathbf y})\sqrt{g}({\mathbf y}))^{-1} \frac{\partial s({\mathbf y})}{\partial y^j} \delta^{(3)}({\mathbf x}-{\mathbf y}), \\
\left[ \gamma_{(r)}({\mathbf x}) , u_j({\mathbf y}) \right] &= -(\rho({\mathbf y})\sqrt{g}({\mathbf y}))^{-1} \frac{\partial \gamma_{(r)}({\mathbf y})}{\partial y^j} \delta^{(3)}({\mathbf x}-{\mathbf y}), \\
\left[ \lambda_{(r)}({\mathbf x}) , u_j({\mathbf y}) \right] &= -(\rho({\mathbf y})\sqrt{g}({\mathbf y}))^{-1} \frac{\partial \lambda_{(r)}({\mathbf y})}{\partial y^j} \delta^{(3)}({\mathbf x}-{\mathbf y}).
\end{align}
From
\begin{align}
\left[ u_i({\mathbf x}) , u_j({\mathbf y}) \right] &= \frac{\partial}{\partial x^i} \left[ \alpha({\mathbf x}) , u_j({\mathbf y}) \right] + \beta({\mathbf x}) \frac{\partial}{\partial x^i} \left[ s({\mathbf x}) , u_j({\mathbf y}) \right] + \frac{\partial s({\mathbf x})}{\partial x^i} \left[ \beta({\mathbf x}) , u_j({\mathbf y}) \right] \nonumber \\
&\quad + \sum_{r=1}^N \left( \gamma_{(r)}({\mathbf x}) \frac{\partial}{\partial x^i} \left[ \lambda_{(r)}({\mathbf x}) , u_j({\mathbf y}) \right] + \frac{\partial \lambda_{(r)}({\mathbf x})}{\partial x^i} \left[ \gamma_{(r)}({\mathbf x}) , u_j({\mathbf y}) \right]  \right), \nonumber \\
&= (\rho({\mathbf y})\sqrt{g}({\mathbf y}))^{-1} \left( \frac{\partial \beta({\mathbf x})}{\partial x^i} \frac{\partial s({\mathbf y})}{\partial y^j} + \sum_{r=1}^N \frac{\partial \gamma_{(r)}({\mathbf x})}{\partial x^i} \frac{\partial \lambda_{(r)}({\mathbf y})}{\partial y^j} \right) \delta^{(3)}({\mathbf x}-{\mathbf y}) \nonumber \\
&\quad -(\rho({\mathbf y})\sqrt{g}({\mathbf y}))^{-1} \left( \frac{\partial \beta({\mathbf y})}{\partial y^j} \frac{\partial s({\mathbf x})}{\partial x^i} + \sum_{r=1}^N \frac{\partial \gamma_{(r)}({\mathbf y})}{\partial y^j} \frac{\partial \lambda_{(r)}({\mathbf x})}{\partial x^i} \right) \delta^{(3)}({\mathbf x}-{\mathbf y}), \nonumber \\
&= (\rho({\mathbf x})\sqrt{g}({\mathbf x}))^{-1} \left( \frac{\partial \beta({\mathbf x})}{\partial x^i} \frac{\partial s({\mathbf x})}{\partial x^j} + \sum_{r=1}^N \frac{\partial \gamma_{(r)}({\mathbf x})}{\partial x^i} \frac{\partial \lambda_{(r)}({\mathbf x})}{\partial x^j} \right) \delta^{(3)}({\mathbf x}-{\mathbf y}) \nonumber \\
&\quad -(\rho({\mathbf x})\sqrt{g}({\mathbf x}))^{-1} \left( \frac{\partial \beta({\mathbf x})}{\partial x^j} \frac{\partial s({\mathbf x})}{\partial x^i} + \sum_{r=1}^N \frac{\partial \gamma_{(r)}({\mathbf x})}{\partial x^j} \frac{\partial \lambda_{(r)}({\mathbf x})}{\partial x^i} \right) \delta^{(3)}({\mathbf x}-{\mathbf y}), \nonumber \\
&= (\sqrt{g}\rho)^{-1} \left( u_{j,i} - u_{i,j} \right) \delta^{(3)}({\mathbf x}-{\mathbf y}), \nonumber \\
&= (\sqrt{g}\rho)^{-1} \left( u_{j:i} - u_{i:j} \right) \delta^{(3)}({\mathbf x}-{\mathbf y}),
\end{align}
one obtains
\begin{align}
\left[ u^i({\mathbf x}) , u^j({\mathbf y}) \right] &= h^{ik}({\mathbf x}) h^{jl}({\mathbf y}) \left[ u_k({\mathbf x}) , u_l({\mathbf y}) \right], \nonumber \\
&= (\sqrt{g}\rho)^{-1} h^{ik}h^{jl} \left( u_{l:k} - u_{k:l} \right) \delta^{(3)}({\mathbf x}-{\mathbf y}), \nonumber \\
&= (\sqrt{g}\rho)^{-1} \left( h^{ik}{u^j}_{:k} - h^{jk}{u^i}_{:k} \right) \delta^{(3)}({\mathbf x}-{\mathbf y}).
\end{align}

\section{Demonstration of Eq.\ \eqref{momi}} \label{amomi}
From Eqs \eqref{hamphys} and \eqref{dup}, one writes
\begin{align}
\left[ u^i({\mathbf x}) , H \right] &= \int d^3y \; \left[ u^i({\mathbf x}) , {\cal H}_D({\mathbf y})\right] - \Gamma^i_{00} -\frac{1}{2} h^{ij} h_{00,j} - h^{ij} g_{jk,0} u^k, \label{fgmn}
\end{align}
where
\begin{align*}
{\cal H}_D=\sqrt{g}\rho\left(\frac{1}{2}g_{kl}u^k u^l +I+\Phi_e\right).
\end{align*}
However,
\begin{align}
\left[ u^i({\mathbf x}) , {\cal H}_D({\mathbf y})\right] &= \sqrt{g} \left[ u^i({\mathbf x}) , \rho({\mathbf y})\right] \left( \frac{1}{2} g_{kl}u^ku^l +I+\Phi_e \right) \nonumber \\
&\quad +\sqrt{g}\rho\left( g_{kl}u^k \left[ u^i({\mathbf x}) , u^l({\mathbf y}) \right] + \left[ u^i({\mathbf x}) , I({\mathbf y}) \right]\right), \nonumber \\
&= \sqrt{g} \left[ u^i({\mathbf x}) , \rho({\mathbf y})\right] \left( \frac{1}{2} g_{kl}u^ku^l +I+\Phi_e \right)+\sqrt{g}\rho\Big( g_{kl}u^k \left[ u^i({\mathbf x}) , u^l({\mathbf y}) \right] \nonumber \\
& \quad + \frac{p}{\rho^2}\left[ u^i({\mathbf x}) , \rho({\mathbf y}) \right] + T\left[ u^i({\mathbf x}) , s({\mathbf y}) \right] \Big).
\end{align}
This means that
\begin{align}
\int d^3y \; \left[ u^i({\mathbf x}) , {\cal H}_D({\mathbf y})\right] &= -h^{ij} \left( \frac{1}{2} g_{kl}u^ku^l +I+\Phi_e \right)_{,j}+g_{kl}u^k\left( h^{ij}{u^l}_{:j} - h^{lj}{u^i}_{:j} \right) \nonumber \\
&\quad  - h^{ij} \left( \frac{p}{\rho}  \right)_{,j} + h^{ij}Ts_{,j}, \nonumber \\
&= -h^{ij} \left( \frac{1}{2} g_{kl}u^ku^l +\Phi_e \right)_{,j}-u^j{u^i}_{:j} + h^{ij}\left(\frac{1}{2} g_{kl}u^ku^l\right)_{:j}-\frac{1}{\rho} h^{ij}p_{,j}, \nonumber \\
&= -h^{ij} \left( \frac{1}{2} g_{kl}u^ku^l +\Phi_e \right)_{,j}-u^j{u^i}_{:j} + h^{ij}\left(\frac{1}{2} g_{\mu\nu}u^\mu u^\nu\right)_{,j} -h^{ij} u_{0:j} \nonumber \\
&\quad -\frac{1}{\rho} h^{ij}p_{,j}, \nonumber \\
&= -h^{ij} \left( \frac{1}{2} g_{kl}u^ku^l +\Phi \right)_{,j}-u^j{u^i}_{:j} + h^{ij}\left(\frac{1}{2} g_{kl}u^k u^l\right)_{,j} -h^{ij} u_{0:j} \nonumber \\
&\quad -\frac{1}{\rho} h^{ij}p_{,j} + h^{ij}h_{00,j} + h^{ij} (g_{0k}u^k)_{,j}, \nonumber \\
&= -u^j{u^i}_{:j}-h^{ij} \left( \frac{1}{\rho}p_{,j}+\Phi_{,j} \right) -h^{ij} u_{0:j} + h^{ij} u_{0,j}, \nonumber \\
&= -u^j{u^i}_{:j}-h^{ij} \left( \frac{1}{\rho}p_{,j}+\Phi_{,j} \right) +h^{ij} \Gamma_{k0j}u^k+h^{ij} \Gamma_{00j}, \nonumber \\
&= -u^j{u^i}_{:j}-h^{ij} \left( \frac{1}{\rho}p_{,j}+\Phi_{,j} \right)+h^{ij} \Gamma_{k0j}u^k + h^{ij} \Gamma_{j0k}u^k -h^{ij} \Gamma_{j0k}u^k \nonumber \\
&\quad +\frac{1}{2}h^{ij} h_{00,j} , \nonumber \\
&= -u^j{u^i}_{:j} -\Gamma^i_{j0}u^j  -h^{ij} \left( \frac{1}{\rho}p_{,j}+\Phi_{,j} \right)+ h^{ij} g_{jk,0}u^k + \frac{1}{2}h^{ij} h_{00,j}, \nonumber \\
&= -u^j{u^i}_{,j} - u^j\Gamma^i_{j\mu}u^\mu -\Gamma^i_{j0}u^j  -h^{ij} \left( \frac{1}{\rho}p_{,j}+\Phi_{,j} \right) \nonumber \\
&\qquad + h^{ij} g_{jk,0}u^k + \frac{1}{2}h^{ij} h_{00,j}, \nonumber \\
&= -u^j{u^i}_{,j} - 2 \Gamma^i_{j0}u^j -  \Gamma^i_{jk}u^ju^k -h^{ij} \left( \frac{1}{\rho}p_{,j}+\Phi_{,j} \right) \nonumber \\ 
&\qquad + h^{ij} g_{jk,0}u^k + \frac{1}{2}h^{ij} h_{00,j}.
\end{align}
Therefore from Eq.\ \eqref{fgmn},
\begin{align}
\left[ u^i({\mathbf x}) , H \right] = -u^j{u^i}_{,j} -\Gamma^i_{00}- 2 \Gamma^i_{j0}u^j -  \Gamma^i_{jk}u^ju^k -h^{ij} \left( \frac{1}{\rho}p_{,j}+\Phi_{,j} \right).
\end{align}
\end{appendices}

\bibliographystyle{ametsoc2014}
\bibliography{refer}

\end{document}